\documentclass[aps,prl,reprint,superscriptaddress,showpacs,showkeys,preprintnumbers,nofootinbib]{revtex4-2}

\usepackage[bookmarks=true]{hyperref}
\usepackage{graphicx,epsfig,color,xcolor}
\usepackage[english]{babel}
\usepackage{amssymb,amsfonts,amsmath}
\usepackage{fixmath}
\usepackage{verbatim}
\usepackage{ulem}

\newcommand{\be}{\begin{equation}} \newcommand{\ee}{\end{equation}}
\newcommand{\bea}{\begin{eqnarray}} \newcommand{\eea}{\end{eqnarray}}
\newcommand{\OO}{\mathcal{O}}
\newcommand{\parent}[1]{\left(#1\right)}
\newcommand{\dd}{\text{d}}
\newcommand{\entropy}{\mathcal{S}}

\newcommand{\newsec}[1]{\textbf{#1.}}

\newcommand{\ene}{\mathcal{E}}

\newcommand{\peq}{\mathcal{P}_{\rm{eq}}}
\newcommand{\plong}{\mathcal{P}_\parallel}
\newcommand{\pperp}{\mathcal{P}_\perp}
\newcommand{\plongh}{\mathcal{P}_\parallel^{\rm{hyd}}}
\newcommand{\pperph}{\mathcal{P}_\perp^{\rm{hyd}}}

\newcommand{\fig}[1]{Fig.~\ref{#1}}

\newcommand{\lp}{\ell_p}

\begin{document}

\title{Cosmic censorship in a (dual) collider}

\author{Marc~Aragonès~Fontboté}
\email{m.aragonesfontbote@students.uu.nl}
\affiliation{Institute for Theoretical Physics, Utrecht University, 3584 CC Utrecht, The Netherlands.}

\author{David~Mateos}
\email{dmateos@fqa.ub.edu}
\affiliation{Departament de Física Quàntica i Astrofísica, Universitat de Barcelona, Martí i Franquès 1,
ES-08028, Barcelona, Spain.}
\affiliation{
Institut de Ciències del Cosmos (ICC), Universitat de Barcelona, Martí i Franquès 1, ES-08028,
Barcelona, Spain.}
\affiliation{
Institució Catalana de Recerca i Estudis Avançats (ICREA), Passeig Lluís Companys 23,
ES-08010, Barcelona, Spain.}

\author{Guillem~Pérez~Martín}
\email{g.perezmartin@students.uu.nl}
\affiliation{Institute for Theoretical Physics, Utrecht University, 3584 CC Utrecht, The Netherlands.}

\author{Wilke~van~der~Schee}
\email{w.vanderschee@uu.nl}
\affiliation{Institute for Theoretical Physics, Utrecht University, 3584 CC Utrecht, The Netherlands.}
\affiliation{
Theoretical Physics Department, CERN CH-1211 Genève 23, Switzerland.
}
\affiliation{Nikhef, Science Park 105, 1098 XG Amsterdam, The Netherlands}

\author{Javier~G.~Subils}
\email{j.gomezsubils@uu.nl}
\affiliation{Institute for Theoretical Physics, Utrecht University, 3584 CC Utrecht, The Netherlands.}

\begin{abstract}
We investigate cosmic censorship in anti-de Sitter space in holographic models in which the ground state is described by a good singularity. These include supersymmetric truncations of string/M-theory, for which a positive-energy theorem holds.  
At the boundary, our solutions describe a boost-invariant fluid in which the  temperature decreases monotonically with time. On the gravity side, they  correspond to black-brane spacetimes with a receding horizon. In classical gravity, curvature invariants at the horizon grow  without bound. In the full theory this regime may or may not be reached. In some cases it is avoided by a phase transition to a regular geometry. In others it is reached but the boundary hydrodynamic evolution can be continued, provided the equation of state at parametrically small energies is known. Both cases require  the inclusion of finite-$N$ or finite-coupling effects.
\end{abstract}
\maketitle
\noindent
\newsec{1.--Introduction}
\label{sec:intro} 
The formation of spacetime singularities is a  prediction of General Relativity (GR) and, at the same time, a signal of its own breakdown.  Mathematically, the term ``singularity'' refers to the incompleteness of the evolution. Here we will adopt a physical definition and think of it as a region of arbitrarily large curvature. From this perspective, the content of the weak cosmic censorship conjecture (WCCC) \cite{Penrose:1969pc} is that, if a singularity forms dynamically from smooth initial data, then it must be hidden behind an event horizon.

Examples of WCCC violation include Choptuik's critical collapse in four dimensions \cite{Choptuik:1992jv} and black-string pinches in the evolution of the Gregory-Laflamme instability in five dimensions \cite{Gregory:1993vy,Lehner:2010pn,Figueras:2022zkg}. Similar pinches appear in the dynamics of five-dimensional black rings \cite{Figueras:2015hkb} and in black hole collisions in higher dimensions \cite{Andrade:2018yqu,Andrade:2019edf,Andrade:2020dgc}. Further plausible violations in four dimensions have also been proposed \cite{Eperon:2019viw}. A common feature of all these examples is the fact that the singularities are localised in a microscopic region of space and time. As a consequence, GR is still expected to provide a reliable description of these systems, with the possible exception of the few quanta resulting from the evaporation of the singularity \cite{Emparan:2020vyf}. Here we will present simple examples in asymptotically anti de-Sitter (AdS) spaces in different dimensions, including $D=4$, in which arbitrarily large curvatures in a macroscopic region are generated upon classical time evolution (see \cite{Horowitz:2016ezu,Crisford:2017zpi} for related examples). 

The essential ingredients are gravity in AdS$_D$ plus a scalar field whose potential is negative and  unbounded from below. This does  not necessarily lead to instabilities. For example, a tachyonic scalar field with negative mass-squared above the Breitenlohner--Freedman (BF) bound \cite{Breitenlohner:1982jf,Breitenlohner:1982bm} is stable. More generally, the existence of a positive-energy theorem is guaranteed provided the potential can be derived from an appropriate superpotential \cite{Townsend:1984iu,Skenderis:1999mm,Amsel:2007im}. Our examples satisfy  this condition. In fact, they include supersymmetric consistent truncations of string/M-theory, for which positivity of the energy follows from  supersymmetry. The embedding in string/M-theory guarantees the existence of a complete dual description in terms of a non-gravitational boundary theory in $d=D-1$ dimensions.

The key idea is to consider models in which the ground-state geometry on the gravity side possesses a ``good singularity'' \cite{Gubser:2000nd}. This geometry can be viewed as the low-temperature limit  of a family of black-brane solutions with regular horizons. Each of these is dual to a static fluid at the boundary in thermal equilibrium. Physically, the singularity does not indicate a pathology of the system but simply that the properties of the ground state, unlike those of thermal states, are not well captured by classical GR.  

Consider now the same fluid undergoing boost-invariant expansion, as in the central region of an ultrarelativistic heavy-ion collision \cite{Bjorken:1982qr} (hence our title). In this case, the energy density dilutes and the fluid temperature decreases monotonically with time. Qualitatively, we expect that the gravitational dual solution should correspond to a black-brane spacetime with a receding horizon. We will see that this is true also quantitatively. After a short initial period out of equilibrium, the fluid becomes well described by hydrodynamics. From this point onward, the gravity solution is well approximated by an adiabatic succession of black-brane geometries with a temperature given by the instantaneous fluid temperature. As the latter decreases, the gravity solution approaches the singular ground state geometry and curvature invariants near the horizon grow without bound. Once they reach the Planck or the string scales, the GR description in the bulk breaks down. At this point, the fluid energy density is parametrically small. In some microscopic models, the boundary hydrodynamic evolution can be continued beyond this point provided the equation of state (EoS) in this limit is known. In others, a phase transition to a regular geometry takes place before the high-curvature regime is reached. In both cases, a correct description requires the inclusion of finite-$N$ or finite-coupling effects, corresponding to quantum gravity or stringy corrections in the bulk, respectively. 

The gravity dual of a conformal, boost-invariant fluid at late times was constructed in \cite{Janik:2005zt}. In this case, conformality implies that the ground state geometry is vacuum AdS and no singularity arises. The construction was generalized to the non-conformal case in \cite{Gursoy:2015nza}. We will see that our solutions approach those in \cite{Gursoy:2015nza} at late times. 

\noindent
\newsec{2.--Holographic model}
\label{sec:setup}
We will present solutions to the following theory:
\begin{equation}\label{eq:actionGH}
S\,=\,\frac{2}{\lp^3}\int \dd^5x\sqrt{-G}\left(\frac{R}{4}-\frac12 \partial_M\phi\partial^M\phi
-V(\phi)\right)\,, 
\end{equation}
where $G_{MN}$ is the bulk metric and  $\lp$ is the Planck length. We assume that the potential follows from a superpotential $W(\phi)$ through the standard relation 
\begin{equation}
    \label{eq:pot.from.superpot}
V(\phi)= -\frac{4}{3}W(\phi)^2 + \frac{1}{2}\left(\frac{\partial W(\phi)}{\partial \phi}\right)^2\,,
\end{equation}
and we take $W(\phi)$ to have  the simple form\footnote{$W$ is of the ``$P_-$''-type  in the language of \cite{Amsel:2007im}. \label{foot}}
\begin{equation}\label{eq:superpotential}
    W(\phi) = \frac{1}{4\gamma^2 L}\Big[ 
    1-6\gamma^2 -\cosh(2\gamma\phi)\Big].
\end{equation}
Here $\gamma$ is a numerical constant and $L$ is the asymptotic AdS radius. The scalar field has mass \mbox{$m^2L^ 2=-3$} around the maximum\footnote{This is outside the double-quantization region $-4\leq m^2L^2 < -3$ \cite{Klebanov:1999tb}.}  of  $V$  at $\phi=0$.  This means that the dual four-dimensional theory is a conformal field theory (CFT) deformed by a dimension $\Delta = 3$ scalar operator. The coefficient of this operator introduces an intrinsic mass scale, $\Lambda$, which breaks conformal invariance. 
The properties of the ground state geometry are determined by the leading exponential in the potential, which for $\phi \to \infty$ behaves as\footnote{Provided $\gamma < 2\gamma_c$ (see Sec.~3).}  
\be
V(\phi)\propto - e^{4\gamma\phi}\,.
\label{run}
\ee
 
This model  captures the essential physics we wish to describe. We focus on $D=5$ for concreteness, but analogous results apply in other dimensions,  including $D=4$.  We choose the exponential  behaviour \eqref{run} because it appears in supersymmetric truncations of string/M-theory to $D=4$ \cite{Duff:1999gh,Faedo:2017fbv,Elander:2020rgv,Cvetic:2000db,Klebanov:2010qs,Dias:2017opt,Gauntlett:2009zw,Gauntlett:2009bh,Donos:2012yu,Duff:1999gh,Hertog:2004rz} and $D=5$ \cite{Freedman:1999gk, Freedman:1999gp,Girardello:1999bd,Petrini:2018pjk,Bobev:2018eer,Pilch:2000fu, Baguet:2015sma,Bena:2010pr,Cassani:2011sv,Buchel:2018bzp,Klebanov:2000hb}, but we expect similar results for other types of unbounded potentials.
In truncations with a single scalar, the latter can be directly identified with $\phi$ in \eqref{eq:pot.from.superpot}. In truncations with multiple scalars, $\phi$  parametrizes a specific direction in field space \cite{Faedo:2017fbv,Elander:2020rgv}. 

\noindent
\newsec{3.--Thermodynamics}
\label{sec:thermo}
The EoS of the boundary theory can be reliably determined from  the black-brane solutions of \eqref{eq:actionGH} provided their curvature obeys 
\begin{equation}
\label{curva}
\lp^2 R \ll 1 \,, \qquad 
\ell_s^2 R \ll 1\,.      
\end{equation}
The Planck length is related to the number of degrees of freedom in the boundary theory through
\begin{equation}
\label{nn}
    N^2=4 \pi^2 L^3/\lp^3 \,,
\end{equation}
meaning that the classical limit in the bulk, $\ell_p \to 0$, corresponds to the planar limit in the boundary theory, $N\to\infty$. The second condition in \eqref{curva} applies in string theory models. In these cases the string length, $\ell_s$, is related to a coupling constant in the boundary theory.

\begin{figure}[t!]
    \centering
    \includegraphics[width=0.47\textwidth]{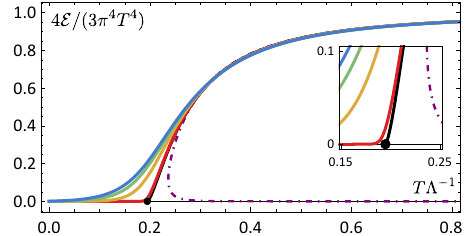}
    \caption{\small  Energy density  as a function of temperature for different choices of $\gamma\leq \gamma_c$, with the same color coding as in Fig.~\ref{fig:EoS}. We also show the case $\gamma=5/10>\gamma_c$ (purple, dot-dashed curve).  The in-set plot zooms into the different $\ene\to 0$ behaviours as $\gamma$ varies from under- to super-critical.}
    \label{fig:three_cases}
\end{figure}

Under assumption \eqref{curva}, the energy density $\ene$ as a function of temperature $T$ takes the form shown in Fig.~\ref{fig:three_cases}. At high temperatures $\ene \propto T^4$, as expected for an ultraviolet (UV)  fixed point. The behaviour at low energy depends  on $\gamma$. If $\gamma < \gamma_c$, with \mbox{$\gamma_c=1/\sqrt{6} \simeq 0.408$},\footnote{In general dimensions $\gamma_c=1/\sqrt{2(D-2)}  $\cite{Gursoy:2010jh}.} then $\ene$ and $T$ approach zero simultaneously. If $\gamma=\gamma_c$ then $T$ approaches a non-zero value as $\ene\to 0$. If $\gamma_c < \gamma\leq  2\gamma_c$ then $T$ diverges as $\ene\to 0$. In all these cases,  the black-brane solutions on the gravity side approach the ground-state geometry  as $\ene\to 0$. We will focus on $\gamma \leq \gamma_c$ because in this case the approach takes place along a thermodynamically stable branch. If \mbox{$\gamma > 2\gamma_c$} the theory is expected to be pathological \cite{Gubser:2000nd}. 

\begin{figure}[h]
    \centering
    \includegraphics[width=0.47\textwidth]{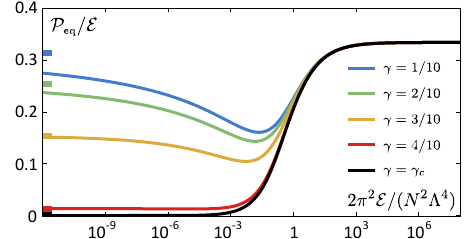}
    \caption{\small Equilibrium pressure  for different  \mbox{$\gamma\in(0,\gamma_c]$}.}
    \label{fig:EoS}
\end{figure}

The EoS, $\peq(\ene)$, is shown in Fig.~\ref{fig:EoS}. 
We see that \mbox{$\peq = \ene/3$} at high energies, as expected for a UV fixed point. The low-energy behaviour is controlled by the leading exponential  \eqref{run} and is given by \mbox{$\peq= (1/3 - 2\gamma^2)\ene$} \cite{Gursoy:2015nza}.  
This value, indicated in Fig.~\ref{fig:EoS} by the small ticks on the left vertical axis, shows  that the IR physics is not conformal. 

The fact that the black-brane solutions approach the singular ground-state geometry as $\ene \to 0$ means that  \eqref{curva} is eventually violated at sufficiently low energy. At this point the EoS above ceases to be valid and a calculation directly in the boundary theory becomes necessary. We will come back to this point below. 

\noindent
\newsec{4.--Time evolution}\footnote{The full numerical code and plot routines can be found  \href{https://www.subils.me/resources/cc-collider/}{\color{blue}{here}}.}
\label{sec:results}
We write the boundary  metric  as
\begin{equation}\label{eq:metric4D}
    \dd s^2_{\text{\tiny B}} = g_{\mu\nu}\dd x^\mu\dd x^\nu = -\dd \tau^2 + \tau^2 \dd y^2  + \dd \vec{x}_\perp^2\,,
\end{equation}
where $\tau$ is the proper time, $y$ is the rapidity and $\vec{x}_\perp$ are the two transverse directions. The fact that this is just the flat Minkowski metric written in Milne coordinates means that no source is turned on for the bulk metric, for which we use the characteristic formulation \cite{Chesler:2009cy, Chesler:2008hg, Chesler:2013lia, vanderSchee:2014qwa, Heller:2012km, Heller:2011ju}
\begin{equation}\label{eq:BI}
    \dd s^2 = -A \dd \tau^2 + 2\dd \tau \dd r + \Sigma^2 \parent{e^{-2B} \dd y^2 + e^{ B}\dd \vec{x}_\perp^2}\,.
\end{equation}
The metric functions $A$, $\Sigma$, $B$ and the scalar $\phi$ depend on $\tau$ and on the holographic direction $r$. After solving the bulk evolution, holographic renormalisation (see \cite{Bianchi:2001de,Bianchi:2001kw,deHaro:2000vlm} and the Supplemental Material (SM)) can be used to read off the boundary stress tensor
\begin{equation}\label{eq:EM.BI}
    \left\langle T^{\mu}_{\ \nu}\right\rangle = \text{diag}\Big( - \ene(\tau),\plong(\tau),\pperp(\tau),\pperp(\tau)\Big)\,,
\end{equation}
whose conservation  implies
\begin{equation}
\label{eq:equationhydro}
    \tau \ene'(\tau) + \ene(\tau) +\plong(\tau) =0\,.
\end{equation} 

\begin{figure}[t]
    \centering
    \includegraphics[width=0.48\textwidth]{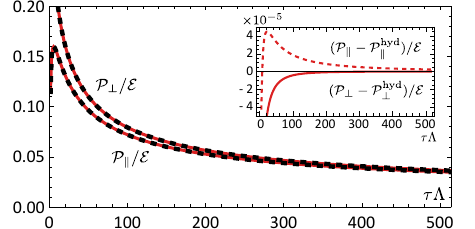}
    \caption{\small Comparison between the exact pressures (solid red) and their hydrodynamic values (dashed black) for $\gamma=4/10$. %
    }
    \label{fig:pressures}
\end{figure}

We set initial conditions at 
$\tau_i = 0.2 \Lambda^{-1}$ such that ${\ene}_i = 75 N^2\Lambda^4/2\pi^2$. We also specify the full radial profiles $B(\tau_i, r)$ and $\phi(\tau_i, r)$ (see \eqref{initial_profiles} in the SM), which fix the initial pressures.  
\begin{figure}[t]
    \centering
    \includegraphics[width=0.48\textwidth]{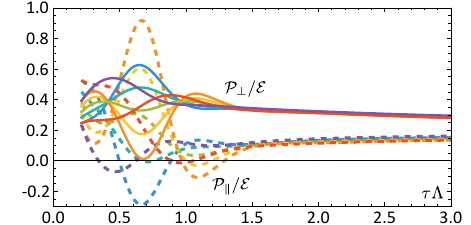}
    \caption{\small Short-time evolution of the  pressures for %
    different choices of initial profiles for $B$ and $\phi$ for $\gamma=4/10$. %
    }
    \label{fig:pressures.initial}
\end{figure}
In Fig.~\ref{fig:pressures.initial} we see that evolutions with different initial conditions quickly converge to one another. As in previous work \cite{Chesler:2009cy,Heller:2011ju}, once this happens the evolution becomes well described by viscous hydrodynamics. This is illustrated by Fig.~\ref{fig:pressures}, 
which compares the exact pressures to their hydrodynamic values including first-order viscous corrections (see \eqref{eq:press_hydro} in the SM). 
We see in the inset panel of Fig.~\ref{fig:pressures} that the relative difference between $\mathcal{P}$ and $\mathcal{P}^{\rm{hyd}}$ is $\lesssim 10^{-4}$ from a very early time. The subsequent evolution can then be determined by solving \eqref{eq:equationhydro} with the replacement $\plong(\tau) \to \plong^{\rm{hyd}}(\tau)$. The  hydrodynamization  at late times can also be understood by comparing the expansion rate of the system with the relaxation rate of perturbations --- see \fig{fig:expansion} in the SM. 

\begin{figure}[t]
    \begin{center}
    \includegraphics[width=0.48\textwidth]{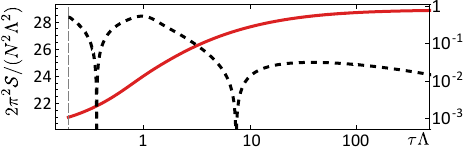}
    \put(-100,23){\footnotesize $
    \left|1-\frac{
    \mbox{Eq.~\eqref{eq.FL}$_{\text{\tiny Left}}$}
    }{
    \mbox{Eq.~\eqref{eq.FL}$_{\text{\tiny Right}}$}
    }
    \right|$}
    \end{center}
    \caption{\small 
    (Solid red) Time evolution of the entropy density per unit rapidity and unit transverse area. 
    (Dashed black) Relative difference between the two sides of Eq.~\eqref{eq.FL}.
    }
    \label{fig:equilibration}
\end{figure}

\begin{figure}[t]
    \centering
    \includegraphics[width=0.48\textwidth]{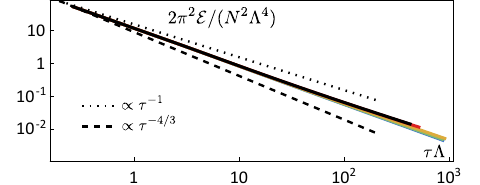}
    \caption{\small Time evolution of the energy density for the same values of $\gamma\leq\gamma_c$ as in Fig.~\ref{fig:EoS}.}
    \label{fig:energydensity}
\end{figure}

In Fig.~\ref{fig:pressures} we  see that $\plong$ and $\pperp$ differ at early times but  converge to one another at late times. At this point, the fluid becomes well described by ideal hydrodynamics and entropy ceases to be produced, as shown in \fig{fig:equilibration}. 
In other words, the fluid has reached local thermal equilibrium and the subsequent evolution can be viewed as adiabatic. We have extracted the entropy density from the area density of the apparent horizon, but we have verified that the latter coincides with the event horizon at late times, as expected in local equilibrium. A further manifestation of this  is the fact that the first law, $\dd E = T\dd S - \mathcal{P} \dd V$, is satisfied. In our case this takes the form 
\begin{equation}
    \label{eq.FL}
    \partial_\tau \Big(\tau \ene(\tau) \Big) = T(\tau)\partial_\tau \entropy(\tau) - \mathcal{P}(\tau)\,,
\end{equation}
where $\entropy$ is the entropy density per unit rapidity and unit transverse area,  $\mathcal{P} \equiv (\plong+2\pperp)/3$ and the time evolution of the energy density is shown in \fig{fig:energydensity}. As we see in \fig{fig:equilibration}, the first law is satisfied with increasing precision as time progresses. 
It is interesting to appreciate the role of each term in \eqref{eq.FL}. The proper volume of a fixed-rapidity region grows linearly with $\tau$. Since the energy density decreases faster than $\tau^{-1}$, the total energy in this region decreases with time, i.e.~$\dd E < 0$, as expected from the dilution caused by the fluid expansion. The total entropy in this region grows monotonically with time and approaches a constant at late times. In this limit $\dd \entropy\to 0$ and the first law is satisfied thanks to the $-\mathcal{P} \dd V$ work term, much like in an expanding FLRW universe. 

\noindent
\newsec{5.--Large curvature}
\label{sec:cosmic}
\begin{figure}[t]
    \centering
    \includegraphics[width=0.48\textwidth]{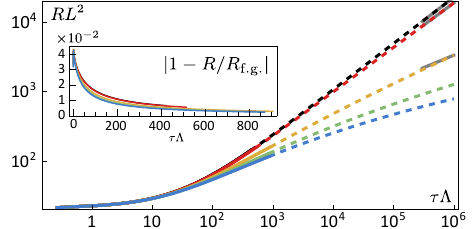}
    \caption{\small Ricci scalar $R$ evaluated at the horizon. Solid curves correspond to the numerical evolutions. Dashed curves correspond to the  fluid/gravity approximation. The limiting behaviour \eqref{eq:limiting_behaviour_curvatures} for $\gamma = 3/10, 4/10$ and $\gamma_c$ is indicated by the grey solid segments on the top-right corner; for the  other two curves this behavior is attained at even later times. The inset panel shows the relative difference between the numerical value of the curvature and its fluid/gravity approximation, $R_{\rm{f.g.}}$, for the times for which they are both available.}
    \label{fig:curvature_invariants}
\end{figure}
In Fig.~\ref{fig:curvature_invariants} we see that the Ricci scalar evaluated at the horizon grows monotonically with time (the same is true for other curvature invariants). As anticipated, the reason is that the expansion of the boundary fluid dilutes the energy density, the bulk horizon recedes deeper into the bulk and the scalar at the horizon ``rolls down'' the unbounded potential. Our numerical simulations, indicated by the solid color curves in Fig.~\ref{fig:curvature_invariants}, confirm this picture. For purely numerical reasons, these simulations become challenging at late times. Fortunately, before this happens the fluid at the boundary has hydrodynamized. From that point onwards, the bulk geometry can be constructed via the fluid/gravity correspondence \cite{Bhattacharyya:2007vjd}. In particular, the curvature at the horizon at a given time $\tau$ is well approximated by its value in the static black-brane solution with energy density $\ene(\tau)$. The inset panel in Fig.~\ref{fig:curvature_invariants} shows that the agreement is better than $1\%$ at the end of the numerical simulations. Using this approximation we can predict the evolution of the curvatures at late times, as indicated by the dashed color curves. In this limit, the behaviour is controlled by the leading exponential \eqref{run}, which implies \cite{Gursoy:2015nza}
\begin{equation}
\label{eq:limiting_behaviour_curvatures}
    R L^2 \propto(\tau \Lambda)^{4\gamma^2}\,.
\end{equation}
This  limit is indicated  by grey solid segments in Fig.~\ref{fig:curvature_invariants}. The agreement with the dashed color curves confirms the accuracy of the fluid/gravity approximation. Note, however, that this assumes that the EoS at parametrically-small energy densities can be obtained from classical GR in the bulk. We will come back to this point in Sec.~7. 

\noindent
\newsec{6.--Sensitivity to initial conditions}
\label{sec:sensi}
In Sec.~4 we considered evolutions with different  initial conditions that respect boost invariance and translation and rotational invariance in the transverse plane. The fact that these quickly converge to one another means that no fine-tuning is necessary in this sector. Consider now perturbations that break boost-invariance locally but preserve it asymptotically. Hydrodynamic modes of this type are expected to be stable and decay as the fluid expands, see e.g.~\cite{Casalderrey-Solana:2020rsj}. In order to check the stability against non-hydrodynamic perturbations we should study their dynamics on top of the boost-invariant plasma. However, the adiabaticity of the late-time evolution suggests that the answer may be captured by the quasi-normal modes  of the static black-brane solutions. We have found no evidence of instabilities in these modes  (see the SM).  In summary, our findings indicate that no fine-tuning is necessary in the sector of initial conditions that preserve boost invariance asymptotically. However, this requirement is itself a form of fine-tuning which, in addition, necessitates  that the total energy in the system  be infinite. 

\noindent
\newsec{7.--Discussion}
\label{sec:discussion}
We have presented asymptotically AdS solutions in which arbitrarily large curvatures are dynamically generated in macroscopic regions  despite the existence of a positive-energy theorem. When condition \eqref{curva} is violated, the evolution in the bulk can no longer be reliably continued using classical GR. We see from \eqref{nn} and \eqref{eq:limiting_behaviour_curvatures} that this happens at a time $\Lambda \tau_p \sim N^{1/3\gamma^2}$. This is longer than (if $\gamma < \gamma_c$) or comparable to (if  $\gamma = \gamma_c$) the time at which the boundary energy density  becomes of order $N^0$, which scales as $\Lambda \tau_0 \sim N^{3/(2-3\gamma^2)}$  (see \eqref{Elate} in the SM). Therefore, finite-$N$ effects must be included in order to assess if the high-curvature regime is reached.  If it is, then the hydrodynamic evolution at the boundary can be continued provided the equation of state can be determined at arbitrarily low energy densities.\footnote{At late times other effects may appear. For instance, $O(N^{-2})$ equilibrium velocity fluctuations introduce power-law tails \cite{Kovtun:2003vj}.}.

In some cases the high-curvature regime is avoided by a phase transition to a regular geometry. This is  illustrated by the eleven-dimensional M-theory model of \cite{Faedo:2017fbv}, whose black brane solutions were constructed in \cite{Elander:2020rgv}. This system can be consistently reduced and truncated to a 
four-dimensional model with six scalars. In this effective description, the horizon values of the scalar fields move along a specific direction in the six-scalar manifold as the energy density decreases. Along this direction, the potential behaves as in \eqref{run} with $\gamma=\gamma_c$. In the $\ene\to 0$ limit, the black-brane solutions approach a supersymmetric geometry that is singular even in eleven dimensions. However, in this limit  there exists another supersymmetric solution that is regular and describes a confining phase.\footnote{More precisely, a gapped phase, see \cite{Faedo:2017fbv}.} The reduction of this solution to four dimensions is singular. Moreover, heating it up does not lead to the appearance of a classical horizon, since its free energy scales as  $N^0$. For these reasons, the confining eleven-dimensional solution is not easily discovered in the four dimensional effective theory. Nevertheless, at times $\tau > \tau_0$ we expect a phase transition, akin to hadronization in Quantum Chromodynamics, from the black-brane solutions to the confining branch. Since the confining solution is regular, this transition cuts off the curvature growth in  eleven dimensions (but not in the effective four-dimensional description).

Our results may be viewed as a rather mild violation of WCCC, because the time it takes for the curvature to become Planckian, $\tau_p$, diverges in the classical limit  $N\to \infty$. It would be interesting to investigate whether instabilities can accelerate the approach to the singularity along the lines of \cite{Bosch:2017ccw,Gursoy:2016ggq}.

\newsec{Acknowledgements}
We thank J.~Casalderrey-Solana, A.~Faedo, U.~G\"ursoy, G.~Horowitz, M.~Sanchez-Garitaonandia and J.~Santos for useful discussions. We are also grateful to R.~Emparan, U.~G\"ursoy and G.~Horowitz for comments on the manuscript. We are especially thankful to R.~Emparan for extensive conversations.  
DM  acknowledges support from grants PID2022-136224NB-C22 and 2021-SGR-872. DM is also  supported by the “Center of Excellence Maria de Maeztu 2020-2023” award to the ICCUB (CEX2019-000918-M) funded by MCIN/AEI/10.13039/501100011033, as well as by the European Research Council (ERC) under the European Union’s Horizon 2020 research and innovation programme (Grant agreement No. 101141909 / project acronym: HoloGW).

\bibliographystyle{apsrev4-1}
\bibliography{main}

\begin{thebibliography}{70}%
\makeatletter
\providecommand \@ifxundefined [1]{%
 \@ifx{#1\undefined}
}%
\providecommand \@ifnum [1]{%
 \ifnum #1\expandafter \@firstoftwo
 \else \expandafter \@secondoftwo
 \fi
}%
\providecommand \@ifx [1]{%
 \ifx #1\expandafter \@firstoftwo
 \else \expandafter \@secondoftwo
 \fi
}%
\providecommand \natexlab [1]{#1}%
\providecommand \enquote  [1]{``#1''}%
\providecommand \bibnamefont  [1]{#1}%
\providecommand \bibfnamefont [1]{#1}%
\providecommand \citenamefont [1]{#1}%
\providecommand \href@noop [0]{\@secondoftwo}%
\providecommand \href [0]{\begingroup \@sanitize@url \@href}%
\providecommand \@href[1]{\@@startlink{#1}\@@href}%
\providecommand \@@href[1]{\endgroup#1\@@endlink}%
\providecommand \@sanitize@url [0]{\catcode `\\12\catcode `\$12\catcode `\&12\catcode `\#12\catcode `\^12\catcode `\_12\catcode `\%12\relax}%
\providecommand \@@startlink[1]{}%
\providecommand \@@endlink[0]{}%
\providecommand \url  [0]{\begingroup\@sanitize@url \@url }%
\providecommand \@url [1]{\endgroup\@href {#1}{\urlprefix }}%
\providecommand \urlprefix  [0]{URL }%
\providecommand \Eprint [0]{\href }%
\providecommand \doibase [0]{http://dx.doi.org/}%
\providecommand \selectlanguage [0]{\@gobble}%
\providecommand \bibinfo  [0]{\@secondoftwo}%
\providecommand \bibfield  [0]{\@secondoftwo}%
\providecommand \translation [1]{[#1]}%
\providecommand \BibitemOpen [0]{}%
\providecommand \bibitemStop [0]{}%
\providecommand \bibitemNoStop [0]{.\EOS\space}%
\providecommand \EOS [0]{\spacefactor3000\relax}%
\providecommand \BibitemShut  [1]{\csname bibitem#1\endcsname}%
\let\auto@bib@innerbib\@empty
\bibitem [{\citenamefont {Penrose}(1969)}]{Penrose:1969pc}%
  \BibitemOpen
  \bibfield  {author} {\bibinfo {author} {\bibfnamefont {R.}~\bibnamefont {Penrose}},\ }\href {\doibase 10.1023/A:1016578408204} {\bibfield  {journal} {\bibinfo  {journal} {Riv. Nuovo Cim.}\ }\textbf {\bibinfo {volume} {1}},\ \bibinfo {pages} {252} (\bibinfo {year} {1969})}\BibitemShut {NoStop}%
\bibitem [{\citenamefont {Choptuik}(1993)}]{Choptuik:1992jv}%
  \BibitemOpen
  \bibfield  {author} {\bibinfo {author} {\bibfnamefont {M.~W.}\ \bibnamefont {Choptuik}},\ }\href {\doibase 10.1103/PhysRevLett.70.9} {\bibfield  {journal} {\bibinfo  {journal} {Phys. Rev. Lett.}\ }\textbf {\bibinfo {volume} {70}},\ \bibinfo {pages} {9} (\bibinfo {year} {1993})}\BibitemShut {NoStop}%
\bibitem [{\citenamefont {Gregory}\ and\ \citenamefont {Laflamme}(1993)}]{Gregory:1993vy}%
  \BibitemOpen
  \bibfield  {author} {\bibinfo {author} {\bibfnamefont {R.}~\bibnamefont {Gregory}}\ and\ \bibinfo {author} {\bibfnamefont {R.}~\bibnamefont {Laflamme}},\ }\href {\doibase 10.1103/PhysRevLett.70.2837} {\bibfield  {journal} {\bibinfo  {journal} {Phys. Rev. Lett.}\ }\textbf {\bibinfo {volume} {70}},\ \bibinfo {pages} {2837} (\bibinfo {year} {1993})},\ \Eprint {http://arxiv.org/abs/hep-th/9301052} {arXiv:hep-th/9301052} \BibitemShut {NoStop}%
\bibitem [{\citenamefont {Lehner}\ and\ \citenamefont {Pretorius}(2010)}]{Lehner:2010pn}%
  \BibitemOpen
  \bibfield  {author} {\bibinfo {author} {\bibfnamefont {L.}~\bibnamefont {Lehner}}\ and\ \bibinfo {author} {\bibfnamefont {F.}~\bibnamefont {Pretorius}},\ }\href {\doibase 10.1103/PhysRevLett.105.101102} {\bibfield  {journal} {\bibinfo  {journal} {Phys. Rev. Lett.}\ }\textbf {\bibinfo {volume} {105}},\ \bibinfo {pages} {101102} (\bibinfo {year} {2010})},\ \Eprint {http://arxiv.org/abs/1006.5960} {arXiv:1006.5960 [hep-th]} \BibitemShut {NoStop}%
\bibitem [{\citenamefont {Figueras}\ \emph {et~al.}(2023)\citenamefont {Figueras}, \citenamefont {Fran\c{c}a}, \citenamefont {Gu},\ and\ \citenamefont {Andrade}}]{Figueras:2022zkg}%
  \BibitemOpen
  \bibfield  {author} {\bibinfo {author} {\bibfnamefont {P.}~\bibnamefont {Figueras}}, \bibinfo {author} {\bibfnamefont {T.}~\bibnamefont {Fran\c{c}a}}, \bibinfo {author} {\bibfnamefont {C.}~\bibnamefont {Gu}}, \ and\ \bibinfo {author} {\bibfnamefont {T.}~\bibnamefont {Andrade}},\ }\href {\doibase 10.1103/PhysRevD.107.044028} {\bibfield  {journal} {\bibinfo  {journal} {Phys. Rev. D}\ }\textbf {\bibinfo {volume} {107}},\ \bibinfo {pages} {044028} (\bibinfo {year} {2023})},\ \Eprint {http://arxiv.org/abs/2210.13501} {arXiv:2210.13501 [hep-th]} \BibitemShut {NoStop}%
\bibitem [{\citenamefont {Figueras}\ \emph {et~al.}(2016)\citenamefont {Figueras}, \citenamefont {Kunesch},\ and\ \citenamefont {Tunyasuvunakool}}]{Figueras:2015hkb}%
  \BibitemOpen
  \bibfield  {author} {\bibinfo {author} {\bibfnamefont {P.}~\bibnamefont {Figueras}}, \bibinfo {author} {\bibfnamefont {M.}~\bibnamefont {Kunesch}}, \ and\ \bibinfo {author} {\bibfnamefont {S.}~\bibnamefont {Tunyasuvunakool}},\ }\href {\doibase 10.1103/PhysRevLett.116.071102} {\bibfield  {journal} {\bibinfo  {journal} {Phys. Rev. Lett.}\ }\textbf {\bibinfo {volume} {116}},\ \bibinfo {pages} {071102} (\bibinfo {year} {2016})},\ \Eprint {http://arxiv.org/abs/1512.04532} {arXiv:1512.04532 [hep-th]} \BibitemShut {NoStop}%
\bibitem [{\citenamefont {Andrade}\ \emph {et~al.}(2019{\natexlab{a}})\citenamefont {Andrade}, \citenamefont {Emparan}, \citenamefont {Licht},\ and\ \citenamefont {Luna}}]{Andrade:2018yqu}%
  \BibitemOpen
  \bibfield  {author} {\bibinfo {author} {\bibfnamefont {T.}~\bibnamefont {Andrade}}, \bibinfo {author} {\bibfnamefont {R.}~\bibnamefont {Emparan}}, \bibinfo {author} {\bibfnamefont {D.}~\bibnamefont {Licht}}, \ and\ \bibinfo {author} {\bibfnamefont {R.}~\bibnamefont {Luna}},\ }\href {\doibase 10.1007/JHEP04(2019)121} {\bibfield  {journal} {\bibinfo  {journal} {JHEP}\ }\textbf {\bibinfo {volume} {04}},\ \bibinfo {pages} {121} (\bibinfo {year} {2019}{\natexlab{a}})},\ \Eprint {http://arxiv.org/abs/1812.05017} {arXiv:1812.05017 [hep-th]} \BibitemShut {NoStop}%
\bibitem [{\citenamefont {Andrade}\ \emph {et~al.}(2019{\natexlab{b}})\citenamefont {Andrade}, \citenamefont {Emparan}, \citenamefont {Licht},\ and\ \citenamefont {Luna}}]{Andrade:2019edf}%
  \BibitemOpen
  \bibfield  {author} {\bibinfo {author} {\bibfnamefont {T.}~\bibnamefont {Andrade}}, \bibinfo {author} {\bibfnamefont {R.}~\bibnamefont {Emparan}}, \bibinfo {author} {\bibfnamefont {D.}~\bibnamefont {Licht}}, \ and\ \bibinfo {author} {\bibfnamefont {R.}~\bibnamefont {Luna}},\ }\href {\doibase 10.1007/JHEP09(2019)099} {\bibfield  {journal} {\bibinfo  {journal} {JHEP}\ }\textbf {\bibinfo {volume} {09}},\ \bibinfo {pages} {099} (\bibinfo {year} {2019}{\natexlab{b}})},\ \Eprint {http://arxiv.org/abs/1908.03424} {arXiv:1908.03424 [hep-th]} \BibitemShut {NoStop}%
\bibitem [{\citenamefont {Andrade}\ \emph {et~al.}(2022)\citenamefont {Andrade}, \citenamefont {Figueras},\ and\ \citenamefont {Sperhake}}]{Andrade:2020dgc}%
  \BibitemOpen
  \bibfield  {author} {\bibinfo {author} {\bibfnamefont {T.}~\bibnamefont {Andrade}}, \bibinfo {author} {\bibfnamefont {P.}~\bibnamefont {Figueras}}, \ and\ \bibinfo {author} {\bibfnamefont {U.}~\bibnamefont {Sperhake}},\ }\href {\doibase 10.1007/JHEP03(2022)111} {\bibfield  {journal} {\bibinfo  {journal} {JHEP}\ }\textbf {\bibinfo {volume} {03}},\ \bibinfo {pages} {111} (\bibinfo {year} {2022})},\ \Eprint {http://arxiv.org/abs/2011.03049} {arXiv:2011.03049 [hep-th]} \BibitemShut {NoStop}%
\bibitem [{\citenamefont {Eperon}\ \emph {et~al.}(2020)\citenamefont {Eperon}, \citenamefont {Ganchev},\ and\ \citenamefont {Santos}}]{Eperon:2019viw}%
  \BibitemOpen
  \bibfield  {author} {\bibinfo {author} {\bibfnamefont {F.~C.}\ \bibnamefont {Eperon}}, \bibinfo {author} {\bibfnamefont {B.}~\bibnamefont {Ganchev}}, \ and\ \bibinfo {author} {\bibfnamefont {J.~E.}\ \bibnamefont {Santos}},\ }\href {\doibase 10.1103/PhysRevD.101.041502} {\bibfield  {journal} {\bibinfo  {journal} {Phys. Rev. D}\ }\textbf {\bibinfo {volume} {101}},\ \bibinfo {pages} {041502} (\bibinfo {year} {2020})},\ \Eprint {http://arxiv.org/abs/1906.11257} {arXiv:1906.11257 [gr-qc]} \BibitemShut {NoStop}%
\bibitem [{\citenamefont {Emparan}(2020)}]{Emparan:2020vyf}%
  \BibitemOpen
  \bibfield  {author} {\bibinfo {author} {\bibfnamefont {R.}~\bibnamefont {Emparan}},\ }\href {\doibase 10.1142/S021827182043021X} {\bibfield  {journal} {\bibinfo  {journal} {Int. J. Mod. Phys. D}\ }\textbf {\bibinfo {volume} {29}},\ \bibinfo {pages} {2043021} (\bibinfo {year} {2020})},\ \Eprint {http://arxiv.org/abs/2005.07389} {arXiv:2005.07389 [hep-th]} \BibitemShut {NoStop}%
\bibitem [{\citenamefont {Horowitz}\ \emph {et~al.}(2016)\citenamefont {Horowitz}, \citenamefont {Santos},\ and\ \citenamefont {Way}}]{Horowitz:2016ezu}%
  \BibitemOpen
  \bibfield  {author} {\bibinfo {author} {\bibfnamefont {G.~T.}\ \bibnamefont {Horowitz}}, \bibinfo {author} {\bibfnamefont {J.~E.}\ \bibnamefont {Santos}}, \ and\ \bibinfo {author} {\bibfnamefont {B.}~\bibnamefont {Way}},\ }\href {\doibase 10.1088/0264-9381/33/19/195007} {\bibfield  {journal} {\bibinfo  {journal} {Class. Quant. Grav.}\ }\textbf {\bibinfo {volume} {33}},\ \bibinfo {pages} {195007} (\bibinfo {year} {2016})},\ \Eprint {http://arxiv.org/abs/1604.06465} {arXiv:1604.06465 [hep-th]} \BibitemShut {NoStop}%
\bibitem [{\citenamefont {Crisford}\ and\ \citenamefont {Santos}(2017)}]{Crisford:2017zpi}%
  \BibitemOpen
  \bibfield  {author} {\bibinfo {author} {\bibfnamefont {T.}~\bibnamefont {Crisford}}\ and\ \bibinfo {author} {\bibfnamefont {J.~E.}\ \bibnamefont {Santos}},\ }\href {\doibase 10.1103/PhysRevLett.118.181101} {\bibfield  {journal} {\bibinfo  {journal} {Phys. Rev. Lett.}\ }\textbf {\bibinfo {volume} {118}},\ \bibinfo {pages} {181101} (\bibinfo {year} {2017})},\ \Eprint {http://arxiv.org/abs/1702.05490} {arXiv:1702.05490 [hep-th]} \BibitemShut {NoStop}%
\bibitem [{\citenamefont {Breitenlohner}\ and\ \citenamefont {Freedman}(1982{\natexlab{a}})}]{Breitenlohner:1982jf}%
  \BibitemOpen
  \bibfield  {author} {\bibinfo {author} {\bibfnamefont {P.}~\bibnamefont {Breitenlohner}}\ and\ \bibinfo {author} {\bibfnamefont {D.~Z.}\ \bibnamefont {Freedman}},\ }\href {\doibase 10.1016/0003-4916(82)90116-6} {\bibfield  {journal} {\bibinfo  {journal} {Annals Phys.}\ }\textbf {\bibinfo {volume} {144}},\ \bibinfo {pages} {249} (\bibinfo {year} {1982}{\natexlab{a}})}\BibitemShut {NoStop}%
\bibitem [{\citenamefont {Breitenlohner}\ and\ \citenamefont {Freedman}(1982{\natexlab{b}})}]{Breitenlohner:1982bm}%
  \BibitemOpen
  \bibfield  {author} {\bibinfo {author} {\bibfnamefont {P.}~\bibnamefont {Breitenlohner}}\ and\ \bibinfo {author} {\bibfnamefont {D.~Z.}\ \bibnamefont {Freedman}},\ }\href {\doibase 10.1016/0370-2693(82)90643-8} {\bibfield  {journal} {\bibinfo  {journal} {Phys. Lett. B}\ }\textbf {\bibinfo {volume} {115}},\ \bibinfo {pages} {197} (\bibinfo {year} {1982}{\natexlab{b}})}\BibitemShut {NoStop}%
\bibitem [{\citenamefont {Townsend}(1984)}]{Townsend:1984iu}%
  \BibitemOpen
  \bibfield  {author} {\bibinfo {author} {\bibfnamefont {P.~K.}\ \bibnamefont {Townsend}},\ }\href {\doibase 10.1016/0370-2693(84)91610-1} {\bibfield  {journal} {\bibinfo  {journal} {Phys. Lett. B}\ }\textbf {\bibinfo {volume} {148}},\ \bibinfo {pages} {55} (\bibinfo {year} {1984})}\BibitemShut {NoStop}%
\bibitem [{\citenamefont {Skenderis}\ and\ \citenamefont {Townsend}(1999)}]{Skenderis:1999mm}%
  \BibitemOpen
  \bibfield  {author} {\bibinfo {author} {\bibfnamefont {K.}~\bibnamefont {Skenderis}}\ and\ \bibinfo {author} {\bibfnamefont {P.~K.}\ \bibnamefont {Townsend}},\ }\href {\doibase 10.1016/S0370-2693(99)01212-5} {\bibfield  {journal} {\bibinfo  {journal} {Phys. Lett. B}\ }\textbf {\bibinfo {volume} {468}},\ \bibinfo {pages} {46} (\bibinfo {year} {1999})},\ \Eprint {http://arxiv.org/abs/hep-th/9909070} {arXiv:hep-th/9909070} \BibitemShut {NoStop}%
\bibitem [{\citenamefont {Amsel}\ \emph {et~al.}(2007)\citenamefont {Amsel}, \citenamefont {Hertog}, \citenamefont {Hollands},\ and\ \citenamefont {Marolf}}]{Amsel:2007im}%
  \BibitemOpen
  \bibfield  {author} {\bibinfo {author} {\bibfnamefont {A.~J.}\ \bibnamefont {Amsel}}, \bibinfo {author} {\bibfnamefont {T.}~\bibnamefont {Hertog}}, \bibinfo {author} {\bibfnamefont {S.}~\bibnamefont {Hollands}}, \ and\ \bibinfo {author} {\bibfnamefont {D.}~\bibnamefont {Marolf}},\ }\href {\doibase 10.1103/PhysRevD.77.049903} {\bibfield  {journal} {\bibinfo  {journal} {Phys. Rev. D}\ }\textbf {\bibinfo {volume} {75}},\ \bibinfo {pages} {084008} (\bibinfo {year} {2007})},\ \bibinfo {note} {[Erratum: Phys.Rev.D 77, 049903 (2008)]},\ \Eprint {http://arxiv.org/abs/hep-th/0701038} {arXiv:hep-th/0701038} \BibitemShut {NoStop}%
\bibitem [{\citenamefont {Gubser}(2000)}]{Gubser:2000nd}%
  \BibitemOpen
  \bibfield  {author} {\bibinfo {author} {\bibfnamefont {S.~S.}\ \bibnamefont {Gubser}},\ }\href {\doibase 10.4310/ATMP.2000.v4.n3.a6} {\bibfield  {journal} {\bibinfo  {journal} {Adv. Theor. Math. Phys.}\ }\textbf {\bibinfo {volume} {4}},\ \bibinfo {pages} {679} (\bibinfo {year} {2000})},\ \Eprint {http://arxiv.org/abs/hep-th/0002160} {arXiv:hep-th/0002160} \BibitemShut {NoStop}%
\bibitem [{\citenamefont {Bjorken}(1983)}]{Bjorken:1982qr}%
  \BibitemOpen
  \bibfield  {author} {\bibinfo {author} {\bibfnamefont {J.~D.}\ \bibnamefont {Bjorken}},\ }\href {\doibase 10.1103/PhysRevD.27.140} {\bibfield  {journal} {\bibinfo  {journal} {Phys. Rev. D}\ }\textbf {\bibinfo {volume} {27}},\ \bibinfo {pages} {140} (\bibinfo {year} {1983})}\BibitemShut {NoStop}%
\bibitem [{\citenamefont {Janik}\ and\ \citenamefont {Peschanski}(2006)}]{Janik:2005zt}%
  \BibitemOpen
  \bibfield  {author} {\bibinfo {author} {\bibfnamefont {R.~A.}\ \bibnamefont {Janik}}\ and\ \bibinfo {author} {\bibfnamefont {R.~B.}\ \bibnamefont {Peschanski}},\ }\href {\doibase 10.1103/PhysRevD.73.045013} {\bibfield  {journal} {\bibinfo  {journal} {Phys. Rev. D}\ }\textbf {\bibinfo {volume} {73}},\ \bibinfo {pages} {045013} (\bibinfo {year} {2006})},\ \Eprint {http://arxiv.org/abs/hep-th/0512162} {arXiv:hep-th/0512162} \BibitemShut {NoStop}%
\bibitem [{\citenamefont {Gursoy}\ \emph {et~al.}(2016)\citenamefont {Gursoy}, \citenamefont {Jarvinen},\ and\ \citenamefont {Policastro}}]{Gursoy:2015nza}%
  \BibitemOpen
  \bibfield  {author} {\bibinfo {author} {\bibfnamefont {U.}~\bibnamefont {Gursoy}}, \bibinfo {author} {\bibfnamefont {M.}~\bibnamefont {Jarvinen}}, \ and\ \bibinfo {author} {\bibfnamefont {G.}~\bibnamefont {Policastro}},\ }\href {\doibase 10.1007/JHEP01(2016)134} {\bibfield  {journal} {\bibinfo  {journal} {JHEP}\ }\textbf {\bibinfo {volume} {01}},\ \bibinfo {pages} {134} (\bibinfo {year} {2016})},\ \Eprint {http://arxiv.org/abs/1507.08628} {arXiv:1507.08628 [hep-th]} \BibitemShut {NoStop}%
\bibitem [{\citenamefont {Klebanov}\ and\ \citenamefont {Witten}(1999)}]{Klebanov:1999tb}%
  \BibitemOpen
  \bibfield  {author} {\bibinfo {author} {\bibfnamefont {I.~R.}\ \bibnamefont {Klebanov}}\ and\ \bibinfo {author} {\bibfnamefont {E.}~\bibnamefont {Witten}},\ }\href {\doibase 10.1016/S0550-3213(99)00387-9} {\bibfield  {journal} {\bibinfo  {journal} {Nucl. Phys. B}\ }\textbf {\bibinfo {volume} {556}},\ \bibinfo {pages} {89} (\bibinfo {year} {1999})},\ \Eprint {http://arxiv.org/abs/hep-th/9905104} {arXiv:hep-th/9905104} \BibitemShut {NoStop}%
\bibitem [{\citenamefont {Duff}\ and\ \citenamefont {Liu}(1999)}]{Duff:1999gh}%
  \BibitemOpen
  \bibfield  {author} {\bibinfo {author} {\bibfnamefont {M.~J.}\ \bibnamefont {Duff}}\ and\ \bibinfo {author} {\bibfnamefont {J.~T.}\ \bibnamefont {Liu}},\ }\href {\doibase 10.1016/S0550-3213(99)00299-0} {\bibfield  {journal} {\bibinfo  {journal} {Nucl. Phys. B}\ }\textbf {\bibinfo {volume} {554}},\ \bibinfo {pages} {237} (\bibinfo {year} {1999})},\ \Eprint {http://arxiv.org/abs/hep-th/9901149} {arXiv:hep-th/9901149} \BibitemShut {NoStop}%
\bibitem [{\citenamefont {Faedo}\ \emph {et~al.}(2017)\citenamefont {Faedo}, \citenamefont {Mateos}, \citenamefont {Pravos},\ and\ \citenamefont {Subils}}]{Faedo:2017fbv}%
  \BibitemOpen
  \bibfield  {author} {\bibinfo {author} {\bibfnamefont {A.~F.}\ \bibnamefont {Faedo}}, \bibinfo {author} {\bibfnamefont {D.}~\bibnamefont {Mateos}}, \bibinfo {author} {\bibfnamefont {D.}~\bibnamefont {Pravos}}, \ and\ \bibinfo {author} {\bibfnamefont {J.~G.}\ \bibnamefont {Subils}},\ }\href {\doibase 10.1007/JHEP06(2017)153} {\bibfield  {journal} {\bibinfo  {journal} {JHEP}\ }\textbf {\bibinfo {volume} {06}},\ \bibinfo {pages} {153} (\bibinfo {year} {2017})},\ \Eprint {http://arxiv.org/abs/1702.05988} {arXiv:1702.05988 [hep-th]} \BibitemShut {NoStop}%
\bibitem [{\citenamefont {Elander}\ \emph {et~al.}(2020)\citenamefont {Elander}, \citenamefont {Faedo}, \citenamefont {Mateos},\ and\ \citenamefont {Subils}}]{Elander:2020rgv}%
  \BibitemOpen
  \bibfield  {author} {\bibinfo {author} {\bibfnamefont {D.}~\bibnamefont {Elander}}, \bibinfo {author} {\bibfnamefont {A.~F.}\ \bibnamefont {Faedo}}, \bibinfo {author} {\bibfnamefont {D.}~\bibnamefont {Mateos}}, \ and\ \bibinfo {author} {\bibfnamefont {J.~G.}\ \bibnamefont {Subils}},\ }\href {\doibase 10.1007/JHEP06(2020)131} {\bibfield  {journal} {\bibinfo  {journal} {JHEP}\ }\textbf {\bibinfo {volume} {06}},\ \bibinfo {pages} {131} (\bibinfo {year} {2020})},\ \Eprint {http://arxiv.org/abs/2002.08279} {arXiv:2002.08279 [hep-th]} \BibitemShut {NoStop}%
\bibitem [{\citenamefont {Cvetic}\ \emph {et~al.}(2003)\citenamefont {Cvetic}, \citenamefont {Gibbons}, \citenamefont {Lu},\ and\ \citenamefont {Pope}}]{Cvetic:2000db}%
  \BibitemOpen
  \bibfield  {author} {\bibinfo {author} {\bibfnamefont {M.}~\bibnamefont {Cvetic}}, \bibinfo {author} {\bibfnamefont {G.~W.}\ \bibnamefont {Gibbons}}, \bibinfo {author} {\bibfnamefont {H.}~\bibnamefont {Lu}}, \ and\ \bibinfo {author} {\bibfnamefont {C.~N.}\ \bibnamefont {Pope}},\ }\href {\doibase 10.1007/s00220-002-0730-3} {\bibfield  {journal} {\bibinfo  {journal} {Commun. Math. Phys.}\ }\textbf {\bibinfo {volume} {232}},\ \bibinfo {pages} {457} (\bibinfo {year} {2003})},\ \Eprint {http://arxiv.org/abs/hep-th/0012011} {arXiv:hep-th/0012011} \BibitemShut {NoStop}%
\bibitem [{\citenamefont {Klebanov}\ and\ \citenamefont {Pufu}(2011)}]{Klebanov:2010qs}%
  \BibitemOpen
  \bibfield  {author} {\bibinfo {author} {\bibfnamefont {I.~R.}\ \bibnamefont {Klebanov}}\ and\ \bibinfo {author} {\bibfnamefont {S.~S.}\ \bibnamefont {Pufu}},\ }\href {\doibase 10.1007/JHEP08(2011)035} {\bibfield  {journal} {\bibinfo  {journal} {JHEP}\ }\textbf {\bibinfo {volume} {08}},\ \bibinfo {pages} {035} (\bibinfo {year} {2011})},\ \Eprint {http://arxiv.org/abs/1006.3587} {arXiv:1006.3587 [hep-th]} \BibitemShut {NoStop}%
\bibitem [{\citenamefont {Dias}\ \emph {et~al.}(2017)\citenamefont {Dias}, \citenamefont {Hartnett}, \citenamefont {Niehoff},\ and\ \citenamefont {Santos}}]{Dias:2017opt}%
  \BibitemOpen
  \bibfield  {author} {\bibinfo {author} {\bibfnamefont {O.~J.~C.}\ \bibnamefont {Dias}}, \bibinfo {author} {\bibfnamefont {G.~S.}\ \bibnamefont {Hartnett}}, \bibinfo {author} {\bibfnamefont {B.~E.}\ \bibnamefont {Niehoff}}, \ and\ \bibinfo {author} {\bibfnamefont {J.~E.}\ \bibnamefont {Santos}},\ }\href {\doibase 10.1007/JHEP11(2017)105} {\bibfield  {journal} {\bibinfo  {journal} {JHEP}\ }\textbf {\bibinfo {volume} {11}},\ \bibinfo {pages} {105} (\bibinfo {year} {2017})},\ \Eprint {http://arxiv.org/abs/1704.02323} {arXiv:1704.02323 [hep-th]} \BibitemShut {NoStop}%
\bibitem [{\citenamefont {Gauntlett}\ \emph {et~al.}(2009)\citenamefont {Gauntlett}, \citenamefont {Kim}, \citenamefont {Varela},\ and\ \citenamefont {Waldram}}]{Gauntlett:2009zw}%
  \BibitemOpen
  \bibfield  {author} {\bibinfo {author} {\bibfnamefont {J.~P.}\ \bibnamefont {Gauntlett}}, \bibinfo {author} {\bibfnamefont {S.}~\bibnamefont {Kim}}, \bibinfo {author} {\bibfnamefont {O.}~\bibnamefont {Varela}}, \ and\ \bibinfo {author} {\bibfnamefont {D.}~\bibnamefont {Waldram}},\ }\href {\doibase 10.1088/1126-6708/2009/04/102} {\bibfield  {journal} {\bibinfo  {journal} {JHEP}\ }\textbf {\bibinfo {volume} {04}},\ \bibinfo {pages} {102} (\bibinfo {year} {2009})},\ \Eprint {http://arxiv.org/abs/0901.0676} {arXiv:0901.0676 [hep-th]} \BibitemShut {NoStop}%
\bibitem [{\citenamefont {Gauntlett}\ \emph {et~al.}(2010)\citenamefont {Gauntlett}, \citenamefont {Sonner},\ and\ \citenamefont {Wiseman}}]{Gauntlett:2009bh}%
  \BibitemOpen
  \bibfield  {author} {\bibinfo {author} {\bibfnamefont {J.~P.}\ \bibnamefont {Gauntlett}}, \bibinfo {author} {\bibfnamefont {J.}~\bibnamefont {Sonner}}, \ and\ \bibinfo {author} {\bibfnamefont {T.}~\bibnamefont {Wiseman}},\ }\href {\doibase 10.1007/JHEP02(2010)060} {\bibfield  {journal} {\bibinfo  {journal} {JHEP}\ }\textbf {\bibinfo {volume} {02}},\ \bibinfo {pages} {060} (\bibinfo {year} {2010})},\ \Eprint {http://arxiv.org/abs/0912.0512} {arXiv:0912.0512 [hep-th]} \BibitemShut {NoStop}%
\bibitem [{\citenamefont {Donos}\ \emph {et~al.}(2013)\citenamefont {Donos}, \citenamefont {Gauntlett}, \citenamefont {Sonner},\ and\ \citenamefont {Withers}}]{Donos:2012yu}%
  \BibitemOpen
  \bibfield  {author} {\bibinfo {author} {\bibfnamefont {A.}~\bibnamefont {Donos}}, \bibinfo {author} {\bibfnamefont {J.~P.}\ \bibnamefont {Gauntlett}}, \bibinfo {author} {\bibfnamefont {J.}~\bibnamefont {Sonner}}, \ and\ \bibinfo {author} {\bibfnamefont {B.}~\bibnamefont {Withers}},\ }\href {\doibase 10.1007/JHEP03(2013)108} {\bibfield  {journal} {\bibinfo  {journal} {JHEP}\ }\textbf {\bibinfo {volume} {03}},\ \bibinfo {pages} {108} (\bibinfo {year} {2013})},\ \Eprint {http://arxiv.org/abs/1212.0871} {arXiv:1212.0871 [hep-th]} \BibitemShut {NoStop}%
\bibitem [{\citenamefont {Hertog}\ and\ \citenamefont {Horowitz}(2004)}]{Hertog:2004rz}%
  \BibitemOpen
  \bibfield  {author} {\bibinfo {author} {\bibfnamefont {T.}~\bibnamefont {Hertog}}\ and\ \bibinfo {author} {\bibfnamefont {G.~T.}\ \bibnamefont {Horowitz}},\ }\href {\doibase 10.1088/1126-6708/2004/07/073} {\bibfield  {journal} {\bibinfo  {journal} {JHEP}\ }\textbf {\bibinfo {volume} {07}},\ \bibinfo {pages} {073} (\bibinfo {year} {2004})},\ \Eprint {http://arxiv.org/abs/hep-th/0406134} {arXiv:hep-th/0406134} \BibitemShut {NoStop}%
\bibitem [{\citenamefont {Freedman}\ \emph {et~al.}(2000)\citenamefont {Freedman}, \citenamefont {Gubser}, \citenamefont {Pilch},\ and\ \citenamefont {Warner}}]{Freedman:1999gk}%
  \BibitemOpen
  \bibfield  {author} {\bibinfo {author} {\bibfnamefont {D.~Z.}\ \bibnamefont {Freedman}}, \bibinfo {author} {\bibfnamefont {S.~S.}\ \bibnamefont {Gubser}}, \bibinfo {author} {\bibfnamefont {K.}~\bibnamefont {Pilch}}, \ and\ \bibinfo {author} {\bibfnamefont {N.~P.}\ \bibnamefont {Warner}},\ }\href {\doibase 10.1088/1126-6708/2000/07/038} {\bibfield  {journal} {\bibinfo  {journal} {JHEP}\ }\textbf {\bibinfo {volume} {07}},\ \bibinfo {pages} {038} (\bibinfo {year} {2000})},\ \Eprint {http://arxiv.org/abs/hep-th/9906194} {arXiv:hep-th/9906194} \BibitemShut {NoStop}%
\bibitem [{\citenamefont {Freedman}\ \emph {et~al.}(1999)\citenamefont {Freedman}, \citenamefont {Gubser}, \citenamefont {Pilch},\ and\ \citenamefont {Warner}}]{Freedman:1999gp}%
  \BibitemOpen
  \bibfield  {author} {\bibinfo {author} {\bibfnamefont {D.~Z.}\ \bibnamefont {Freedman}}, \bibinfo {author} {\bibfnamefont {S.~S.}\ \bibnamefont {Gubser}}, \bibinfo {author} {\bibfnamefont {K.}~\bibnamefont {Pilch}}, \ and\ \bibinfo {author} {\bibfnamefont {N.~P.}\ \bibnamefont {Warner}},\ }\href {\doibase 10.4310/ATMP.1999.v3.n2.a7} {\bibfield  {journal} {\bibinfo  {journal} {Adv. Theor. Math. Phys.}\ }\textbf {\bibinfo {volume} {3}},\ \bibinfo {pages} {363} (\bibinfo {year} {1999})},\ \Eprint {http://arxiv.org/abs/hep-th/9904017} {arXiv:hep-th/9904017} \BibitemShut {NoStop}%
\bibitem [{\citenamefont {Girardello}\ \emph {et~al.}(2000)\citenamefont {Girardello}, \citenamefont {Petrini}, \citenamefont {Porrati},\ and\ \citenamefont {Zaffaroni}}]{Girardello:1999bd}%
  \BibitemOpen
  \bibfield  {author} {\bibinfo {author} {\bibfnamefont {L.}~\bibnamefont {Girardello}}, \bibinfo {author} {\bibfnamefont {M.}~\bibnamefont {Petrini}}, \bibinfo {author} {\bibfnamefont {M.}~\bibnamefont {Porrati}}, \ and\ \bibinfo {author} {\bibfnamefont {A.}~\bibnamefont {Zaffaroni}},\ }\href {\doibase 10.1016/S0550-3213(99)00764-6} {\bibfield  {journal} {\bibinfo  {journal} {Nucl. Phys. B}\ }\textbf {\bibinfo {volume} {569}},\ \bibinfo {pages} {451} (\bibinfo {year} {2000})},\ \Eprint {http://arxiv.org/abs/hep-th/9909047} {arXiv:hep-th/9909047} \BibitemShut {NoStop}%
\bibitem [{\citenamefont {Petrini}\ \emph {et~al.}(2018)\citenamefont {Petrini}, \citenamefont {Samtleben}, \citenamefont {Schmidt},\ and\ \citenamefont {Skenderis}}]{Petrini:2018pjk}%
  \BibitemOpen
  \bibfield  {author} {\bibinfo {author} {\bibfnamefont {M.}~\bibnamefont {Petrini}}, \bibinfo {author} {\bibfnamefont {H.}~\bibnamefont {Samtleben}}, \bibinfo {author} {\bibfnamefont {S.}~\bibnamefont {Schmidt}}, \ and\ \bibinfo {author} {\bibfnamefont {K.}~\bibnamefont {Skenderis}},\ }\href {\doibase 10.1007/JHEP07(2018)026} {\bibfield  {journal} {\bibinfo  {journal} {JHEP}\ }\textbf {\bibinfo {volume} {07}},\ \bibinfo {pages} {026} (\bibinfo {year} {2018})},\ \Eprint {http://arxiv.org/abs/1805.01919} {arXiv:1805.01919 [hep-th]} \BibitemShut {NoStop}%
\bibitem [{\citenamefont {Bobev}\ \emph {et~al.}(2018)\citenamefont {Bobev}, \citenamefont {Gautason}, \citenamefont {Niehoff},\ and\ \citenamefont {van Muiden}}]{Bobev:2018eer}%
  \BibitemOpen
  \bibfield  {author} {\bibinfo {author} {\bibfnamefont {N.}~\bibnamefont {Bobev}}, \bibinfo {author} {\bibfnamefont {F.~F.}\ \bibnamefont {Gautason}}, \bibinfo {author} {\bibfnamefont {B.~E.}\ \bibnamefont {Niehoff}}, \ and\ \bibinfo {author} {\bibfnamefont {J.}~\bibnamefont {van Muiden}},\ }\href {\doibase 10.1007/JHEP10(2018)058} {\bibfield  {journal} {\bibinfo  {journal} {JHEP}\ }\textbf {\bibinfo {volume} {10}},\ \bibinfo {pages} {058} (\bibinfo {year} {2018})},\ \Eprint {http://arxiv.org/abs/1805.03623} {arXiv:1805.03623 [hep-th]} \BibitemShut {NoStop}%
\bibitem [{\citenamefont {Pilch}\ and\ \citenamefont {Warner}(2002)}]{Pilch:2000fu}%
  \BibitemOpen
  \bibfield  {author} {\bibinfo {author} {\bibfnamefont {K.}~\bibnamefont {Pilch}}\ and\ \bibinfo {author} {\bibfnamefont {N.~P.}\ \bibnamefont {Warner}},\ }\href {\doibase 10.4310/ATMP.2000.v4.n3.a5} {\bibfield  {journal} {\bibinfo  {journal} {Adv. Theor. Math. Phys.}\ }\textbf {\bibinfo {volume} {4}},\ \bibinfo {pages} {627} (\bibinfo {year} {2002})},\ \Eprint {http://arxiv.org/abs/hep-th/0006066} {arXiv:hep-th/0006066} \BibitemShut {NoStop}%
\bibitem [{\citenamefont {Baguet}\ \emph {et~al.}(2015)\citenamefont {Baguet}, \citenamefont {Hohm},\ and\ \citenamefont {Samtleben}}]{Baguet:2015sma}%
  \BibitemOpen
  \bibfield  {author} {\bibinfo {author} {\bibfnamefont {A.}~\bibnamefont {Baguet}}, \bibinfo {author} {\bibfnamefont {O.}~\bibnamefont {Hohm}}, \ and\ \bibinfo {author} {\bibfnamefont {H.}~\bibnamefont {Samtleben}},\ }\href {\doibase 10.1103/PhysRevD.92.065004} {\bibfield  {journal} {\bibinfo  {journal} {Phys. Rev. D}\ }\textbf {\bibinfo {volume} {92}},\ \bibinfo {pages} {065004} (\bibinfo {year} {2015})},\ \Eprint {http://arxiv.org/abs/1506.01385} {arXiv:1506.01385 [hep-th]} \BibitemShut {NoStop}%
\bibitem [{\citenamefont {Bena}\ \emph {et~al.}(2011)\citenamefont {Bena}, \citenamefont {Giecold}, \citenamefont {Grana}, \citenamefont {Halmagyi},\ and\ \citenamefont {Orsi}}]{Bena:2010pr}%
  \BibitemOpen
  \bibfield  {author} {\bibinfo {author} {\bibfnamefont {I.}~\bibnamefont {Bena}}, \bibinfo {author} {\bibfnamefont {G.}~\bibnamefont {Giecold}}, \bibinfo {author} {\bibfnamefont {M.}~\bibnamefont {Grana}}, \bibinfo {author} {\bibfnamefont {N.}~\bibnamefont {Halmagyi}}, \ and\ \bibinfo {author} {\bibfnamefont {F.}~\bibnamefont {Orsi}},\ }\href {\doibase 10.1007/JHEP04(2011)021} {\bibfield  {journal} {\bibinfo  {journal} {JHEP}\ }\textbf {\bibinfo {volume} {04}},\ \bibinfo {pages} {021} (\bibinfo {year} {2011})},\ \Eprint {http://arxiv.org/abs/1008.0983} {arXiv:1008.0983 [hep-th]} \BibitemShut {NoStop}%
\bibitem [{\citenamefont {Cassani}\ and\ \citenamefont {Faedo}(2011)}]{Cassani:2011sv}%
  \BibitemOpen
  \bibfield  {author} {\bibinfo {author} {\bibfnamefont {D.}~\bibnamefont {Cassani}}\ and\ \bibinfo {author} {\bibfnamefont {A.~F.}\ \bibnamefont {Faedo}},\ }\href {\doibase 10.1007/JHEP05(2011)013} {\bibfield  {journal} {\bibinfo  {journal} {JHEP}\ }\textbf {\bibinfo {volume} {05}},\ \bibinfo {pages} {013} (\bibinfo {year} {2011})},\ \Eprint {http://arxiv.org/abs/1102.5344} {arXiv:1102.5344 [hep-th]} \BibitemShut {NoStop}%
\bibitem [{\citenamefont {Buchel}(2019)}]{Buchel:2018bzp}%
  \BibitemOpen
  \bibfield  {author} {\bibinfo {author} {\bibfnamefont {A.}~\bibnamefont {Buchel}},\ }\href {\doibase 10.1007/JHEP01(2019)207} {\bibfield  {journal} {\bibinfo  {journal} {JHEP}\ }\textbf {\bibinfo {volume} {01}},\ \bibinfo {pages} {207} (\bibinfo {year} {2019})},\ \Eprint {http://arxiv.org/abs/1809.08484} {arXiv:1809.08484 [hep-th]} \BibitemShut {NoStop}%
\bibitem [{\citenamefont {Klebanov}\ and\ \citenamefont {Strassler}(2000)}]{Klebanov:2000hb}%
  \BibitemOpen
  \bibfield  {author} {\bibinfo {author} {\bibfnamefont {I.~R.}\ \bibnamefont {Klebanov}}\ and\ \bibinfo {author} {\bibfnamefont {M.~J.}\ \bibnamefont {Strassler}},\ }\href {\doibase 10.1088/1126-6708/2000/08/052} {\bibfield  {journal} {\bibinfo  {journal} {JHEP}\ }\textbf {\bibinfo {volume} {08}},\ \bibinfo {pages} {052} (\bibinfo {year} {2000})},\ \Eprint {http://arxiv.org/abs/hep-th/0007191} {arXiv:hep-th/0007191} \BibitemShut {NoStop}%
\bibitem [{\citenamefont {Gursoy}(2011)}]{Gursoy:2010jh}%
  \BibitemOpen
  \bibfield  {author} {\bibinfo {author} {\bibfnamefont {U.}~\bibnamefont {Gursoy}},\ }\href {\doibase 10.1007/JHEP01(2011)086} {\bibfield  {journal} {\bibinfo  {journal} {JHEP}\ }\textbf {\bibinfo {volume} {01}},\ \bibinfo {pages} {086} (\bibinfo {year} {2011})},\ \Eprint {http://arxiv.org/abs/1007.0500} {arXiv:1007.0500 [hep-th]} \BibitemShut {NoStop}%
\bibitem [{\citenamefont {Chesler}\ and\ \citenamefont {Yaffe}(2010)}]{Chesler:2009cy}%
  \BibitemOpen
  \bibfield  {author} {\bibinfo {author} {\bibfnamefont {P.~M.}\ \bibnamefont {Chesler}}\ and\ \bibinfo {author} {\bibfnamefont {L.~G.}\ \bibnamefont {Yaffe}},\ }\href {\doibase 10.1103/PhysRevD.82.026006} {\bibfield  {journal} {\bibinfo  {journal} {Phys. Rev. D}\ }\textbf {\bibinfo {volume} {82}},\ \bibinfo {pages} {026006} (\bibinfo {year} {2010})},\ \Eprint {http://arxiv.org/abs/0906.4426} {arXiv:0906.4426 [hep-th]} \BibitemShut {NoStop}%
\bibitem [{\citenamefont {Chesler}\ and\ \citenamefont {Yaffe}(2009)}]{Chesler:2008hg}%
  \BibitemOpen
  \bibfield  {author} {\bibinfo {author} {\bibfnamefont {P.~M.}\ \bibnamefont {Chesler}}\ and\ \bibinfo {author} {\bibfnamefont {L.~G.}\ \bibnamefont {Yaffe}},\ }\href {\doibase 10.1103/PhysRevLett.102.211601} {\bibfield  {journal} {\bibinfo  {journal} {Phys. Rev. Lett.}\ }\textbf {\bibinfo {volume} {102}},\ \bibinfo {pages} {211601} (\bibinfo {year} {2009})},\ \Eprint {http://arxiv.org/abs/0812.2053} {arXiv:0812.2053 [hep-th]} \BibitemShut {NoStop}%
\bibitem [{\citenamefont {Chesler}\ and\ \citenamefont {Yaffe}(2014)}]{Chesler:2013lia}%
  \BibitemOpen
  \bibfield  {author} {\bibinfo {author} {\bibfnamefont {P.~M.}\ \bibnamefont {Chesler}}\ and\ \bibinfo {author} {\bibfnamefont {L.~G.}\ \bibnamefont {Yaffe}},\ }\href {\doibase 10.1007/JHEP07(2014)086} {\bibfield  {journal} {\bibinfo  {journal} {JHEP}\ }\textbf {\bibinfo {volume} {07}},\ \bibinfo {pages} {086} (\bibinfo {year} {2014})},\ \Eprint {http://arxiv.org/abs/1309.1439} {arXiv:1309.1439 [hep-th]} \BibitemShut {NoStop}%
\bibitem [{\citenamefont {van~der Schee}(2014)}]{vanderSchee:2014qwa}%
  \BibitemOpen
  \bibfield  {author} {\bibinfo {author} {\bibfnamefont {W.}~\bibnamefont {van~der Schee}},\ }\href@noop {} {\  (\bibinfo {year} {2014})},\ \Eprint {http://arxiv.org/abs/1407.1849} {arXiv:1407.1849 [hep-th]} \BibitemShut {NoStop}%
\bibitem [{\citenamefont {Heller}\ \emph {et~al.}(2012{\natexlab{a}})\citenamefont {Heller}, \citenamefont {Mateos}, \citenamefont {van~der Schee},\ and\ \citenamefont {Trancanelli}}]{Heller:2012km}%
  \BibitemOpen
  \bibfield  {author} {\bibinfo {author} {\bibfnamefont {M.~P.}\ \bibnamefont {Heller}}, \bibinfo {author} {\bibfnamefont {D.}~\bibnamefont {Mateos}}, \bibinfo {author} {\bibfnamefont {W.}~\bibnamefont {van~der Schee}}, \ and\ \bibinfo {author} {\bibfnamefont {D.}~\bibnamefont {Trancanelli}},\ }\href {\doibase 10.1103/PhysRevLett.108.191601} {\bibfield  {journal} {\bibinfo  {journal} {Phys. Rev. Lett.}\ }\textbf {\bibinfo {volume} {108}},\ \bibinfo {pages} {191601} (\bibinfo {year} {2012}{\natexlab{a}})},\ \Eprint {http://arxiv.org/abs/1202.0981} {arXiv:1202.0981 [hep-th]} \BibitemShut {NoStop}%
\bibitem [{\citenamefont {Heller}\ \emph {et~al.}(2012{\natexlab{b}})\citenamefont {Heller}, \citenamefont {Janik},\ and\ \citenamefont {Witaszczyk}}]{Heller:2011ju}%
  \BibitemOpen
  \bibfield  {author} {\bibinfo {author} {\bibfnamefont {M.~P.}\ \bibnamefont {Heller}}, \bibinfo {author} {\bibfnamefont {R.~A.}\ \bibnamefont {Janik}}, \ and\ \bibinfo {author} {\bibfnamefont {P.}~\bibnamefont {Witaszczyk}},\ }\href {\doibase 10.1103/PhysRevLett.108.201602} {\bibfield  {journal} {\bibinfo  {journal} {Phys. Rev. Lett.}\ }\textbf {\bibinfo {volume} {108}},\ \bibinfo {pages} {201602} (\bibinfo {year} {2012}{\natexlab{b}})},\ \Eprint {http://arxiv.org/abs/1103.3452} {arXiv:1103.3452 [hep-th]} \BibitemShut {NoStop}%
\bibitem [{\citenamefont {Bianchi}\ \emph {et~al.}(2001)\citenamefont {Bianchi}, \citenamefont {Freedman},\ and\ \citenamefont {Skenderis}}]{Bianchi:2001de}%
  \BibitemOpen
  \bibfield  {author} {\bibinfo {author} {\bibfnamefont {M.}~\bibnamefont {Bianchi}}, \bibinfo {author} {\bibfnamefont {D.~Z.}\ \bibnamefont {Freedman}}, \ and\ \bibinfo {author} {\bibfnamefont {K.}~\bibnamefont {Skenderis}},\ }\href {\doibase 10.1088/1126-6708/2001/08/041} {\bibfield  {journal} {\bibinfo  {journal} {JHEP}\ }\textbf {\bibinfo {volume} {08}},\ \bibinfo {pages} {041} (\bibinfo {year} {2001})},\ \Eprint {http://arxiv.org/abs/hep-th/0105276} {arXiv:hep-th/0105276} \BibitemShut {NoStop}%
\bibitem [{\citenamefont {Bianchi}\ \emph {et~al.}(2002)\citenamefont {Bianchi}, \citenamefont {Freedman},\ and\ \citenamefont {Skenderis}}]{Bianchi:2001kw}%
  \BibitemOpen
  \bibfield  {author} {\bibinfo {author} {\bibfnamefont {M.}~\bibnamefont {Bianchi}}, \bibinfo {author} {\bibfnamefont {D.~Z.}\ \bibnamefont {Freedman}}, \ and\ \bibinfo {author} {\bibfnamefont {K.}~\bibnamefont {Skenderis}},\ }\href {\doibase 10.1016/S0550-3213(02)00179-7} {\bibfield  {journal} {\bibinfo  {journal} {Nucl. Phys. B}\ }\textbf {\bibinfo {volume} {631}},\ \bibinfo {pages} {159} (\bibinfo {year} {2002})},\ \Eprint {http://arxiv.org/abs/hep-th/0112119} {arXiv:hep-th/0112119} \BibitemShut {NoStop}%
\bibitem [{\citenamefont {de~Haro}\ \emph {et~al.}(2001)\citenamefont {de~Haro}, \citenamefont {Solodukhin},\ and\ \citenamefont {Skenderis}}]{deHaro:2000vlm}%
  \BibitemOpen
  \bibfield  {author} {\bibinfo {author} {\bibfnamefont {S.}~\bibnamefont {de~Haro}}, \bibinfo {author} {\bibfnamefont {S.~N.}\ \bibnamefont {Solodukhin}}, \ and\ \bibinfo {author} {\bibfnamefont {K.}~\bibnamefont {Skenderis}},\ }\href {\doibase 10.1007/s002200100381} {\bibfield  {journal} {\bibinfo  {journal} {Commun. Math. Phys.}\ }\textbf {\bibinfo {volume} {217}},\ \bibinfo {pages} {595} (\bibinfo {year} {2001})},\ \Eprint {http://arxiv.org/abs/hep-th/0002230} {arXiv:hep-th/0002230} \BibitemShut {NoStop}%
\bibitem [{\citenamefont {Bhattacharyya}\ \emph {et~al.}(2008)\citenamefont {Bhattacharyya}, \citenamefont {Hubeny}, \citenamefont {Minwalla},\ and\ \citenamefont {Rangamani}}]{Bhattacharyya:2007vjd}%
  \BibitemOpen
  \bibfield  {author} {\bibinfo {author} {\bibfnamefont {S.}~\bibnamefont {Bhattacharyya}}, \bibinfo {author} {\bibfnamefont {V.~E.}\ \bibnamefont {Hubeny}}, \bibinfo {author} {\bibfnamefont {S.}~\bibnamefont {Minwalla}}, \ and\ \bibinfo {author} {\bibfnamefont {M.}~\bibnamefont {Rangamani}},\ }\href {\doibase 10.1088/1126-6708/2008/02/045} {\bibfield  {journal} {\bibinfo  {journal} {JHEP}\ }\textbf {\bibinfo {volume} {02}},\ \bibinfo {pages} {045} (\bibinfo {year} {2008})},\ \Eprint {http://arxiv.org/abs/0712.2456} {arXiv:0712.2456 [hep-th]} \BibitemShut {NoStop}%
\bibitem [{\citenamefont {Casalderrey-Solana}\ \emph {et~al.}(2021)\citenamefont {Casalderrey-Solana}, \citenamefont {Milhano}, \citenamefont {Pablos}, \citenamefont {Rajagopal},\ and\ \citenamefont {Yao}}]{Casalderrey-Solana:2020rsj}%
  \BibitemOpen
  \bibfield  {author} {\bibinfo {author} {\bibfnamefont {J.}~\bibnamefont {Casalderrey-Solana}}, \bibinfo {author} {\bibfnamefont {J.~G.}\ \bibnamefont {Milhano}}, \bibinfo {author} {\bibfnamefont {D.}~\bibnamefont {Pablos}}, \bibinfo {author} {\bibfnamefont {K.}~\bibnamefont {Rajagopal}}, \ and\ \bibinfo {author} {\bibfnamefont {X.}~\bibnamefont {Yao}},\ }\href {\doibase 10.1007/JHEP05(2021)230} {\bibfield  {journal} {\bibinfo  {journal} {JHEP}\ }\textbf {\bibinfo {volume} {05}},\ \bibinfo {pages} {230} (\bibinfo {year} {2021})},\ \Eprint {http://arxiv.org/abs/2010.01140} {arXiv:2010.01140 [hep-ph]} \BibitemShut {NoStop}%
\bibitem [{\citenamefont {Kovtun}\ and\ \citenamefont {Yaffe}(2003)}]{Kovtun:2003vj}%
  \BibitemOpen
  \bibfield  {author} {\bibinfo {author} {\bibfnamefont {P.}~\bibnamefont {Kovtun}}\ and\ \bibinfo {author} {\bibfnamefont {L.~G.}\ \bibnamefont {Yaffe}},\ }\href {\doibase 10.1103/PhysRevD.68.025007} {\bibfield  {journal} {\bibinfo  {journal} {Phys. Rev. D}\ }\textbf {\bibinfo {volume} {68}},\ \bibinfo {pages} {025007} (\bibinfo {year} {2003})},\ \Eprint {http://arxiv.org/abs/hep-th/0303010} {arXiv:hep-th/0303010} \BibitemShut {NoStop}%
\bibitem [{\citenamefont {Bosch}\ \emph {et~al.}(2017)\citenamefont {Bosch}, \citenamefont {Buchel},\ and\ \citenamefont {Lehner}}]{Bosch:2017ccw}%
  \BibitemOpen
  \bibfield  {author} {\bibinfo {author} {\bibfnamefont {P.}~\bibnamefont {Bosch}}, \bibinfo {author} {\bibfnamefont {A.}~\bibnamefont {Buchel}}, \ and\ \bibinfo {author} {\bibfnamefont {L.}~\bibnamefont {Lehner}},\ }\href {\doibase 10.1007/JHEP07(2017)135} {\bibfield  {journal} {\bibinfo  {journal} {JHEP}\ }\textbf {\bibinfo {volume} {07}},\ \bibinfo {pages} {135} (\bibinfo {year} {2017})},\ \Eprint {http://arxiv.org/abs/1704.05454} {arXiv:1704.05454 [hep-th]} \BibitemShut {NoStop}%
\bibitem [{\citenamefont {G\"ursoy}\ \emph {et~al.}(2016)\citenamefont {G\"ursoy}, \citenamefont {Jansen},\ and\ \citenamefont {van~der Schee}}]{Gursoy:2016ggq}%
  \BibitemOpen
  \bibfield  {author} {\bibinfo {author} {\bibfnamefont {U.}~\bibnamefont {G\"ursoy}}, \bibinfo {author} {\bibfnamefont {A.}~\bibnamefont {Jansen}}, \ and\ \bibinfo {author} {\bibfnamefont {W.}~\bibnamefont {van~der Schee}},\ }\href {\doibase 10.1103/PhysRevD.94.061901} {\bibfield  {journal} {\bibinfo  {journal} {Phys. Rev. D}\ }\textbf {\bibinfo {volume} {94}},\ \bibinfo {pages} {061901} (\bibinfo {year} {2016})},\ \Eprint {http://arxiv.org/abs/1603.07724} {arXiv:1603.07724 [hep-th]} \BibitemShut {NoStop}%
\bibitem [{\citenamefont {Bea}\ \emph {et~al.}(2021)\citenamefont {Bea}, \citenamefont {Dias}, \citenamefont {Giannakopoulos}, \citenamefont {Mateos}, \citenamefont {Sanchez-Garitaonandia}, \citenamefont {Santos},\ and\ \citenamefont {Zilhao}}]{Bea:2020ees}%
  \BibitemOpen
  \bibfield  {author} {\bibinfo {author} {\bibfnamefont {Y.}~\bibnamefont {Bea}}, \bibinfo {author} {\bibfnamefont {O.~J.~C.}\ \bibnamefont {Dias}}, \bibinfo {author} {\bibfnamefont {T.}~\bibnamefont {Giannakopoulos}}, \bibinfo {author} {\bibfnamefont {D.}~\bibnamefont {Mateos}}, \bibinfo {author} {\bibfnamefont {M.}~\bibnamefont {Sanchez-Garitaonandia}}, \bibinfo {author} {\bibfnamefont {J.~E.}\ \bibnamefont {Santos}}, \ and\ \bibinfo {author} {\bibfnamefont {M.}~\bibnamefont {Zilhao}},\ }\href {\doibase 10.1007/JHEP02(2021)061} {\bibfield  {journal} {\bibinfo  {journal} {JHEP}\ }\textbf {\bibinfo {volume} {02}},\ \bibinfo {pages} {061} (\bibinfo {year} {2021})},\ \Eprint {http://arxiv.org/abs/2007.06467} {arXiv:2007.06467 [hep-th]} \BibitemShut {NoStop}%
\bibitem [{\citenamefont {Policastro}\ \emph {et~al.}(2001)\citenamefont {Policastro}, \citenamefont {Son},\ and\ \citenamefont {Starinets}}]{Policastro:2001yc}%
  \BibitemOpen
  \bibfield  {author} {\bibinfo {author} {\bibfnamefont {G.}~\bibnamefont {Policastro}}, \bibinfo {author} {\bibfnamefont {D.~T.}\ \bibnamefont {Son}}, \ and\ \bibinfo {author} {\bibfnamefont {A.~O.}\ \bibnamefont {Starinets}},\ }\href {\doibase 10.1103/PhysRevLett.87.081601} {\bibfield  {journal} {\bibinfo  {journal} {Phys. Rev. Lett.}\ }\textbf {\bibinfo {volume} {87}},\ \bibinfo {pages} {081601} (\bibinfo {year} {2001})},\ \Eprint {http://arxiv.org/abs/hep-th/0104066} {arXiv:hep-th/0104066} \BibitemShut {NoStop}%
\bibitem [{\citenamefont {Eling}\ and\ \citenamefont {Oz}(2011)}]{Eling:2011ms}%
  \BibitemOpen
  \bibfield  {author} {\bibinfo {author} {\bibfnamefont {C.}~\bibnamefont {Eling}}\ and\ \bibinfo {author} {\bibfnamefont {Y.}~\bibnamefont {Oz}},\ }\href {\doibase 10.1007/JHEP06(2011)007} {\bibfield  {journal} {\bibinfo  {journal} {JHEP}\ }\textbf {\bibinfo {volume} {06}},\ \bibinfo {pages} {007} (\bibinfo {year} {2011})},\ \Eprint {http://arxiv.org/abs/1103.1657} {arXiv:1103.1657 [hep-th]} \BibitemShut {NoStop}%
\bibitem [{\citenamefont {Baier}\ \emph {et~al.}(2008)\citenamefont {Baier}, \citenamefont {Romatschke}, \citenamefont {Son}, \citenamefont {Starinets},\ and\ \citenamefont {Stephanov}}]{Baier:2007ix}%
  \BibitemOpen
  \bibfield  {author} {\bibinfo {author} {\bibfnamefont {R.}~\bibnamefont {Baier}}, \bibinfo {author} {\bibfnamefont {P.}~\bibnamefont {Romatschke}}, \bibinfo {author} {\bibfnamefont {D.~T.}\ \bibnamefont {Son}}, \bibinfo {author} {\bibfnamefont {A.~O.}\ \bibnamefont {Starinets}}, \ and\ \bibinfo {author} {\bibfnamefont {M.~A.}\ \bibnamefont {Stephanov}},\ }\href {\doibase 10.1088/1126-6708/2008/04/100} {\bibfield  {journal} {\bibinfo  {journal} {JHEP}\ }\textbf {\bibinfo {volume} {04}},\ \bibinfo {pages} {100} (\bibinfo {year} {2008})},\ \Eprint {http://arxiv.org/abs/0712.2451} {arXiv:0712.2451 [hep-th]} \BibitemShut {NoStop}%
\bibitem [{\citenamefont {Jansen}(2017)}]{Jansen:2017oag}%
  \BibitemOpen
  \bibfield  {author} {\bibinfo {author} {\bibfnamefont {A.}~\bibnamefont {Jansen}},\ }\href {\doibase 10.1140/epjp/i2017-11825-9} {\bibfield  {journal} {\bibinfo  {journal} {Eur. Phys. J. Plus}\ }\textbf {\bibinfo {volume} {132}},\ \bibinfo {pages} {546} (\bibinfo {year} {2017})},\ \Eprint {http://arxiv.org/abs/1709.09178} {arXiv:1709.09178 [gr-qc]} \BibitemShut {NoStop}%
\bibitem [{\citenamefont {Kaminski}\ \emph {et~al.}(2010)\citenamefont {Kaminski}, \citenamefont {Landsteiner}, \citenamefont {Mas}, \citenamefont {Shock},\ and\ \citenamefont {Tarrio}}]{Kaminski:2009dh}%
  \BibitemOpen
  \bibfield  {author} {\bibinfo {author} {\bibfnamefont {M.}~\bibnamefont {Kaminski}}, \bibinfo {author} {\bibfnamefont {K.}~\bibnamefont {Landsteiner}}, \bibinfo {author} {\bibfnamefont {J.}~\bibnamefont {Mas}}, \bibinfo {author} {\bibfnamefont {J.~P.}\ \bibnamefont {Shock}}, \ and\ \bibinfo {author} {\bibfnamefont {J.}~\bibnamefont {Tarrio}},\ }\href {\doibase 10.1007/JHEP02(2010)021} {\bibfield  {journal} {\bibinfo  {journal} {JHEP}\ }\textbf {\bibinfo {volume} {02}},\ \bibinfo {pages} {021} (\bibinfo {year} {2010})},\ \Eprint {http://arxiv.org/abs/0911.3610} {arXiv:0911.3610 [hep-th]} \BibitemShut {NoStop}%
\bibitem [{\citenamefont {Betzios}\ \emph {et~al.}(2018)\citenamefont {Betzios}, \citenamefont {G\"ursoy}, \citenamefont {J\"arvinen},\ and\ \citenamefont {Policastro}}]{Betzios:2017dol}%
  \BibitemOpen
  \bibfield  {author} {\bibinfo {author} {\bibfnamefont {P.}~\bibnamefont {Betzios}}, \bibinfo {author} {\bibfnamefont {U.}~\bibnamefont {G\"ursoy}}, \bibinfo {author} {\bibfnamefont {M.}~\bibnamefont {J\"arvinen}}, \ and\ \bibinfo {author} {\bibfnamefont {G.}~\bibnamefont {Policastro}},\ }\href {\doibase 10.1103/PhysRevD.97.081901} {\bibfield  {journal} {\bibinfo  {journal} {Phys. Rev. D}\ }\textbf {\bibinfo {volume} {97}},\ \bibinfo {pages} {081901} (\bibinfo {year} {2018})},\ \Eprint {http://arxiv.org/abs/1708.02252} {arXiv:1708.02252 [hep-th]} \BibitemShut {NoStop}%
\bibitem [{\citenamefont {Betzios}\ \emph {et~al.}(2020)\citenamefont {Betzios}, \citenamefont {G\"ursoy}, \citenamefont {J\"arvinen},\ and\ \citenamefont {Policastro}}]{Betzios:2018kwn}%
  \BibitemOpen
  \bibfield  {author} {\bibinfo {author} {\bibfnamefont {P.}~\bibnamefont {Betzios}}, \bibinfo {author} {\bibfnamefont {U.}~\bibnamefont {G\"ursoy}}, \bibinfo {author} {\bibfnamefont {M.}~\bibnamefont {J\"arvinen}}, \ and\ \bibinfo {author} {\bibfnamefont {G.}~\bibnamefont {Policastro}},\ }\href {\doibase 10.1103/PhysRevD.101.086026} {\bibfield  {journal} {\bibinfo  {journal} {Phys. Rev. D}\ }\textbf {\bibinfo {volume} {101}},\ \bibinfo {pages} {086026} (\bibinfo {year} {2020})},\ \Eprint {http://arxiv.org/abs/1807.01718} {arXiv:1807.01718 [hep-th]} \BibitemShut {NoStop}%
\bibitem [{\citenamefont {Grozdanov}\ \emph {et~al.}(2019)\citenamefont {Grozdanov}, \citenamefont {Kovtun}, \citenamefont {Starinets},\ and\ \citenamefont {Tadi\'c}}]{Grozdanov:2019uhi}%
  \BibitemOpen
  \bibfield  {author} {\bibinfo {author} {\bibfnamefont {S.}~\bibnamefont {Grozdanov}}, \bibinfo {author} {\bibfnamefont {P.~K.}\ \bibnamefont {Kovtun}}, \bibinfo {author} {\bibfnamefont {A.~O.}\ \bibnamefont {Starinets}}, \ and\ \bibinfo {author} {\bibfnamefont {P.}~\bibnamefont {Tadi\'c}},\ }\href {\doibase 10.1007/JHEP11(2019)097} {\bibfield  {journal} {\bibinfo  {journal} {JHEP}\ }\textbf {\bibinfo {volume} {11}},\ \bibinfo {pages} {097} (\bibinfo {year} {2019})},\ \Eprint {http://arxiv.org/abs/1904.12862} {arXiv:1904.12862 [hep-th]} \BibitemShut {NoStop}%
\bibitem [{\citenamefont {Kovtun}\ and\ \citenamefont {Starinets}(2005)}]{Kovtun:2005ev}%
  \BibitemOpen
  \bibfield  {author} {\bibinfo {author} {\bibfnamefont {P.~K.}\ \bibnamefont {Kovtun}}\ and\ \bibinfo {author} {\bibfnamefont {A.~O.}\ \bibnamefont {Starinets}},\ }\href {\doibase 10.1103/PhysRevD.72.086009} {\bibfield  {journal} {\bibinfo  {journal} {Phys. Rev. D}\ }\textbf {\bibinfo {volume} {72}},\ \bibinfo {pages} {086009} (\bibinfo {year} {2005})},\ \Eprint {http://arxiv.org/abs/hep-th/0506184} {arXiv:hep-th/0506184} \BibitemShut {NoStop}%
\bibitem [{\citenamefont {Ecker}\ \emph {et~al.}(2022)\citenamefont {Ecker}, \citenamefont {van~der Schee}, \citenamefont {Mateos},\ and\ \citenamefont {Casalderrey-Solana}}]{Ecker:2021cvz}%
  \BibitemOpen
  \bibfield  {author} {\bibinfo {author} {\bibfnamefont {C.}~\bibnamefont {Ecker}}, \bibinfo {author} {\bibfnamefont {W.}~\bibnamefont {van~der Schee}}, \bibinfo {author} {\bibfnamefont {D.}~\bibnamefont {Mateos}}, \ and\ \bibinfo {author} {\bibfnamefont {J.}~\bibnamefont {Casalderrey-Solana}},\ }\href {\doibase 10.1007/JHEP03(2022)137} {\bibfield  {journal} {\bibinfo  {journal} {JHEP}\ }\textbf {\bibinfo {volume} {03}},\ \bibinfo {pages} {137} (\bibinfo {year} {2022})},\ \Eprint {http://arxiv.org/abs/2109.10355} {arXiv:2109.10355 [hep-th]} \BibitemShut {NoStop}%
\end{thebibliography}%

\onecolumngrid
\appendix

\section{SUPPLEMENTAL MATERIAL}
\subsection{Holographic renormalisation}
Holography gives access to the microscopic evolution of expectation values of operators of strongly coupled quantum field theories in terms of boundary data through the well-known holographic renormalization procedure from Refs.~\cite{Bianchi:2001de,Bianchi:2001kw,deHaro:2000vlm}. For completeness and ease of reference, let us collect the relevant formulae for our system in our conventions.

Imposing that the four-dimensional boundary metric is conformal to that in \eqref{eq:metric4D} implies  the following near-boundary behaviour of the  five-dimensional metric functions and scalar field:
\begin{equation}\label{eq:UV.BI}
    \begin{array}{ll}
    \displaystyle
    A(\tau,r) = \frac{r^2}{L^2}+2 \xi (\tau ) r+\cdots+\frac{a_4(\tau
   )}{r^2} + \cdots \,,
   &\displaystyle \qquad
   B(\tau,r) = -\frac{2}{3}\log (\tau)+\cdots+\frac{b_4(\tau
   )}{r^4}+ \cdots\,,
   \\[3mm] \displaystyle
   \Sigma(\tau,r) = \frac{\tau^ {\frac{1}{3}} r}{L} + \cdots\,,
   & \displaystyle \qquad \phi(\tau,r) = \frac{\phi _1}{r}+\cdots+\frac{\phi _3(\tau
   )}{r^3}+\cdots\,.
    \end{array}
\end{equation}
Here we show explicitly only the leading-order terms, as well as the first appearance of the coefficients that are not determined by the asymptotic expansion of the equations. The constant $\phi_1$ corresponds to the source of the operator dual to $\phi$, and sets the energy scale $\Lambda = \phi_1/L^2$ of the explicit breaking of conformal symmetry in the UV. We use this scale to set the units in the plots. Moreover, $\xi(\tau)$ is a remaining gauge freedom that we use to fix the position of the apparent horizon. Finally, $a_4(\tau)$, $b_4(\tau)$, $\phi_3(\tau)$ are related to the energy density, pressures and vacuum expectation value of the operator dual to $\phi$ through  holographic renormalization, as we discuss next.

Holographic renormalization is better implemented in Fefferman-Graham (FG) coordinates coordinates. In these coordinates the asymptotic boundary is at $z = 0$ and $G_{zz} = L^2/z^2$. Near this region, the metric reads
\begin{equation}\label{eq:FGansatz}
    \dd s^ 2 = \frac{L^ 2}{z^2}\parent{\dd z^ 2 + g_{\mu\nu}\dd x^\mu\dd x^\nu}\,,
\end{equation}
with 
\begin{equation}
    g_{\mu\nu} = g_{\mu\nu}^{(0)}(t,\vec x) + g_{\mu\nu}^{(2)}(t,\vec x) z^ 2 + g_{\mu\nu}^{(4)}(t,\vec x) z^ 4 + \cdots\,;
\end{equation}
and the scalar decaying as
\begin{equation}
    \phi = \phi^{(1)}(t,\vec x)  z +     \phi^{(3)}(t,\vec x) z^ 3 +\cdots\,.
\end{equation}
Higher-order terms in the expansion are determined in terms of these ones. As indicated, $g_{\mu\nu}^{(i)}$ and $\phi^{(i)}$ generically depend on all the gauge theory coordinates. For our purposes, $g_{\mu\nu}^{(0)}$ will be fixed to the flat Minkowski metric written in convenient coordinates, while $\phi^{(1)}(t,\vec x) = \Lambda$ is constant and identified as the source of the relevant operator that triggers the non-trivial renormalization group flow in  the boundary dual theory. Note that $g_{\mu\nu}^{(2)} = -\Lambda^2 L^2 g_{\mu\nu}^{(0)}/3$ is also fixed by the equations, leaving $g_{\mu\nu}^{(4)}(t,\vec{x})$ and $\phi^{(3)}(t,\vec{x})$ as the only undetermined functions in the previous expansions.

The expressions from Refs.~\cite{Bianchi:2001de,Bianchi:2001kw,deHaro:2000vlm} simplify enormously in our case because the boundary metric is conformally flat (in fact, flat) and the source of the scalar is constant. The vacuum expectation value of the energy-momentum tensor and of the operator dual to $\phi$ take the form (see Ref.~\cite{Bea:2020ees} for a similar model)
\begin{equation}\label{eq:EMtensor}
    \left\langle T_{\mu \nu}\right\rangle = \frac{2L^3}{\lp^3}\parent{g_{\mu\nu}^{(4)}(t,\vec{x}) + g_{\mu\nu}^{(0)}(t,\vec{x})\parent{\Lambda \phi^{(3)}(t,\vec{x}) - \frac{\Lambda^ 4}{18}+\lambda_4 \Lambda^4}}\,,\quad \langle\OO_\phi \rangle = 
    \frac{2L^3}{\lp^3}\parent{
    {-2\phi^{(3)}(t,\vec{x})-4\lambda_4 \Lambda^ 3}
    }\,.
\end{equation}
Different choices of $\lambda_4$ in these expressions correspond to different renormalization schemes. In supersymmetric models, the supersymmetry-preserving scheme corresponds to setting $\lambda_4$ equal to the coefficient of the 
$\phi^4$-term in the expansion of the superpotential around $\phi=0$. In this scheme the energy density of the ground state vanishes exactly. We adopt this choice here and thus set  $\lambda_4=-\gamma^2/6$. 

\subsection{Static solutions}
In the main text we mentioned that the thermodynamics of the boundary theory can be determined from the static black brane solutions on the gravity side. A standard way to construct these solutions is to impose homogeneity and isotropy with respect to the boundary directions in Eq.~\eqref{eq:FGansatz}.
This means that $g_{\mu\nu}^{(0)} = \eta_{\mu\nu} = \text{diag}(-1,1,1,1)$, and the metric simplifies to
\begin{equation}\label{eq:static}
    \dd s^ 2 = \frac{L^ 2}{z^2}\parent{-f(z)\dd t^ 2 + g(z)\dd \vec{x}^ 2 + \dd z^ 2}\,.
\end{equation}
Regularity at the horizon $z=z_H$ requires that the fields behave as\footnote{Note that $f$ has a double pole at the horizon because we work in coordinates in which $G_{zz}$ is finite at the horizon.}
\begin{equation}
    f(z) = f_H (z-z_H)^2+ \cdots\,,\qquad
    g(z) = g_H +\cdots\,,\qquad
    \phi(z) = \phi_H+ \cdots\,,
\end{equation}
where the dots stand for higher-order terms in an expansion in powers of $(z-z_H)$. 
The requirement that \mbox{$f(0) = g(0) = 1$} fixes the coefficients $f_H$ and $g_H$, in such a way that in the end we are left with a one-parameter familiy of solutions labeled by $\phi_H$, which we find numerically. From the fall-off of the fields at the boundary we can read off the energy density $\ene$ and the pressure $\mathcal{P}$ of equilibrium states using 
Eq.~\eqref{eq:EMtensor}. In contrast, the entropy density and the temperature are given in terms of horizon data through
\begin{equation}
s = \frac{2\pi L^3}{\lp^3}\frac{g_H^{3/2}}{z_H^3}\,,\qquad T = \frac{f_H^{1/2}}{2\pi}\,.
\end{equation}

\subsection{Boost-invariant solutions}
We will now specify to the boost-invariant case. Here $g_{\mu\nu}^{(0)} = \gamma_{\mu\nu} = \text{diag}(-1,\tau^2,1,1)$ is the flat space metric in boost-invariant coordinates, see Eq.~\eqref{eq:metric4D}. By relating the FG coordinates to the Eddington-Finkelstein coordinates used in \eqref{eq:BI}, we can write the energy-momentum tensor \eqref{eq:EMtensor} in terms of boundary data as
\begin{align}
\label{eq:fullEvolution}
\begin{split}
   \ene(\tau) &= \frac{\Lambda^4N^2}{2\pi^2}\times\left[-\frac{3 a_4(\tau )}{4\Lambda^4 L^6} +\frac{1}{4} \frac{\xi
   (\tau )^2}{\Lambda^2}-\frac{\phi _3(\tau )}{\Lambda^3L^6}+\frac{7}{36}-\lambda _4\right] \,,
   \\[2mm]
   \plong(\tau) &= \frac{\Lambda^4N^2}{2\pi^2}\times\frac{1}{108\tau^4}\Bigg[\tau^4 \left(108 \lambda_4-5\right) 
+ \frac{32 \tau^2 + 48 \tau^3 \xi(\tau) - 9 \tau^4 \xi(\tau)^2}{\Lambda^2} 
\\&+ \frac{36 + 72 \tau \xi(\tau) + 54 \tau^2 \xi(\tau)^2 + 18 \tau^3 \xi(\tau)^3 
- 27 \tau^4 \left( a_4(\tau)L^{-6} + 8 b_4(\tau)L^{-8}\right)}{\Lambda^4} 
+ \frac{36 \tau^4 \phi_3(\tau)}{L^6 \Lambda^3}
\Bigg]\,,
   \\[2mm]
  \pperp(\tau) &= \frac{\Lambda^4N^2}{2\pi^2}\times\frac{1}{108 \tau ^4}\Bigg[
\tau^4 \left( 108 \lambda_4-5\right) 
- \frac{16 \tau^2 + 24 \tau^3 \xi(\tau) + 9 \tau^4 \xi(\tau)^2}{\Lambda^2} \\
&- \frac{18 + 36 \tau \xi(\tau) + 27 \tau^2 \xi(\tau)^2 + 9 \tau^3 \xi(\tau)^3 
+ 27 \tau^4 \left(a_4(\tau)L^{-6} - 4 b_4(\tau)L^{-8}\right)}{\Lambda^4} 
+ \frac{36 \tau^4 \phi_3(\tau)}{L^6 \Lambda^3}\Bigg]\,.
\end{split}
\end{align}

\subsection{Initial conditions}
To initialise a simulation, it is necessary to provide the initial value of  $a_4(\tau_i)$ at the time slice $\tau=\tau_i$, as well as the full bulk profiles of $B(\tau_i,r)$ and $\phi(\tau_i,r)$ in a given gauge specified by $\xi(\tau_i)$. In this paper we choose the initial profiles 
\begin{equation}\label{initial_profiles}
    \begin{aligned}
    B(\tau_i,r)&= -\frac{2}{3}\log(\tau_i)-\frac{2 L^2}{3\tau_ir}+\frac{L^4}{3\tau_i^2r^2}-\frac{2L^6(3+2\Lambda^2\tau_i^2)}{27\tau_i^3r^3}+\frac{L^8\cos(3\Lambda L^2/r)}{r^4}(b_4(\tau_i) - 1/(6\tau_i^4)) \,,
   \\[2mm]
   \phi(\tau_i,r) &= \Lambda L^2/r + \phi_3(\tau_i)L^6/r^3 \,,
    \end{aligned}
\end{equation}
in the gauge where $\partial_r A = 2r$ at $r=\infty$, which corresponds to  $\xi(\tau_i) = 0$). In this way, every choice of $a_4(\tau_i)$, $\phi_3(\tau_i)$ and $b_3(\tau_i)$ leads to a different choice of the initial energy density and pressures.

\subsection{Transport and hydrodynamics}
\begin{figure}[t]
    \centering
    \includegraphics[width=0.49\textwidth]{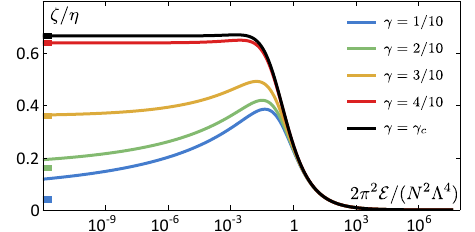}
    \caption{\small Ratio of viscosities  for different choices of $\gamma\in(0,\gamma_c]$.}
    \label{fig:viscosities}
\end{figure}

In addition to the thermodynamics, from the static solutions we can also extract the shear viscosity $\eta$ and the bulk viscosity $\zeta$. As usual, $\eta=s/4\pi$ \cite{Policastro:2001yc}, with $s$ the entropy density per unit volume. We use \cite{Eling:2011ms} to compute $\zeta/\eta$, which we show in Fig.~\ref{fig:viscosities}. We see that at high energies $\zeta = 0$, as expected for a UV fixed point. The low-energy behaviour is controlled by the leading exponential  \eqref{run} and is given by \cite{Gubser:2000nd,Gursoy:2015nza} $\zeta/\eta = 4\gamma^2$. These values, indicated in Fig.~\ref{fig:viscosities} by the small ticks on the left vertical axis, show  that the IR physics is not conformal. 

The hydrodynamic pressures including first-order viscosity corrections are given by \cite{Baier:2007ix} 
\begin{equation}\label{eq:press_hydro}
    \plongh = \peq-\frac{1}{3 \tau}(4 \eta + 3 \zeta )\,,\qquad  \pperph = \peq +\frac{1}{3 \tau}(2 \eta -3 \zeta )\,,
\end{equation}
where $\peq, \eta$ and $\zeta$ are the functions of $\ene$ plotted in Figs.~\ref{fig:EoS} and \ref{fig:viscosities}.

\begin{figure}[t]   
    \centering
    \includegraphics[width=.96\textwidth]{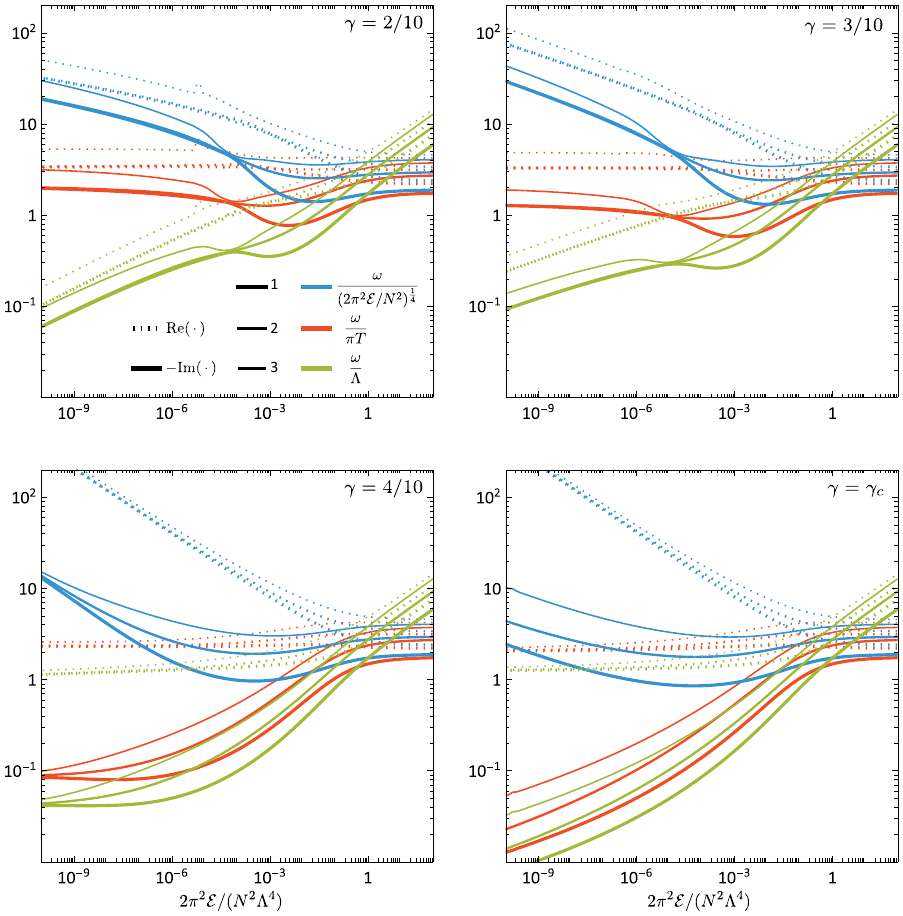}
    \caption{\small Real (dotted) and imaginary (solid) part of the first three quasinormal modes (thick, normal, thin) for $\gamma = 2/10$ (top) and $4/10$ (bottom), normalised by the energy density (blue), the temperature (red) and the intrinsic scale $\Lambda$ (green), as a function of $\ene/\Lambda$. %
    }
    \label{fig:QNMs_zero_momentum}
\end{figure}
\begin{figure}[t!]
    \centering
    \includegraphics[width=0.45\textwidth]{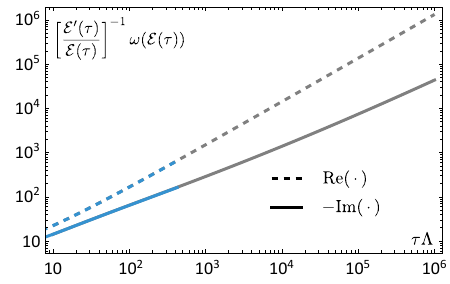}\hfill
    \caption{\small Ratio between the relaxation rate and expansion rate for $\gamma=\gamma_c$. Blue curves stand for the times where the full evolution in the bulk is available, while the grey curves correspond to their extension to later times using the hydrodynamic approximation.}
    \label{fig:expansion}
\end{figure}

\subsection{Quasi-normal modes}
Quasi-normal modes (QNM) characterise the way in which the plasma, or equivalently the black branes solutions, relax after suffering a small perturbation. We computed them following Ref.~\cite{Jansen:2017oag} and their notation. First the static solutions in Eq.\eqref{eq:static} are perturbed through 
\begin{equation}
    G_{MN}\mapsto G_{MN} + \delta G_{MN} \,,\quad \phi \mapsto \phi+ \delta \phi\,.
\end{equation}
We look for plane-wave solutions and thus we decompose the perturbations as $\delta G_{MN} = e^{-i(\omega t - q x)}h_{MN}$ and \mbox{$\delta \phi = e^{-i(\omega t - q x)}\varphi$}. This leads to a system of second-order, linear, homogeneous differential equations for $h_{MN}$ and $ \varphi$. These have some redundancies which can be removed by defining four gauge-invariant fields: one helicity-two field $Z_3$, one helicity-one field $Z_1$ and two coupled scalar fields $Z_2$ and  $Z_\phi$. Their equations take the schematic form
\begin{equation}
    A_\alpha[\omega,q](z) Z_\alpha''(z) +
    B_\alpha[\omega,q](z) Z_\alpha'(z)+
    C_\alpha
[\omega,q](u) Z_\alpha(z)=0\,,\quad \alpha = \{1,3\}\,.
\end{equation}
and
\begin{equation}\label{eq:scalrChannel}
     \mathbf{A}[\omega,q](z) \left(\begin{array}{c}
         Z_2''(z)  \\
         Z_\phi''(z)
    \end{array}\right) +
    \mathbf{B}[\omega,q](z)  \left(\begin{array}{c}
         Z_2''(z)  \\
         Z_\phi''(z)
    \end{array}\right)
    + \mathbf{C}[\omega,q](z) 
    \left(\begin{array}{c}
         Z_2''(z)  \\
         Z_\phi''(z)
    \end{array}\right)\,.
\end{equation}
The coefficients depend on $\omega$ and $q$, as well as on the radial coordinate through the background functions of the metric and the scalar. Now the problem reduces to finding the eigenvalues of the differential operators. Following Ref.~\cite{Jansen:2017oag}, we do this by means of  pseudo-spectral methods (see Ref.~\cite{Kaminski:2009dh} for an alternative approach).

We first examine the case with zero spatial momentum, $q=0$. In this case the equation for $Z_1$ coincides with the equation for $Z_3$ and one of the equations in the scalar channel. Therefore, it is enough to solve Eq.~\eqref{eq:scalrChannel}. The most relevant result for this paper is that we have found no QNM that develops a positive imaginary part.  The frequencies for the three  QNMs with the smallest  imaginary parts are shown in \fig{fig:QNMs_zero_momentum} for different values of $\gamma$. 
At high energies we see the expected conformal behavior $\omega\propto T \propto \ene^{1/4}$. At low energies, the two natural scales given by $T$ and $\ene^{1/4}$ become hierarchically separated, because the temperature  in this limit behaves as 
\begin{equation}
\label{Tlate}
    T \propto \ene^{\frac{1-6\gamma^2}{4-6\gamma^2}} \ll \ene^{\frac{1}{4}} \,.
\end{equation}
This raises the question of which of these two scales sets the frequency of the QNMs as $\ene \to 0$. Fig.~\eqref{fig:QNMs_zero_momentum} indicates that it is the temperature. Indeed, we see that the blue curves diverge in this limit, whereas the red curves seem to saturate. Note, however, that while the real part of the QNMs saturates to an $O(1)$-constant in units of the temperature as $\gamma$ approaches its critical value, the imaginary part becomes smaller and smaller. This is consistent with the behaviour seen for $\gamma=\gamma_c$ in \cite{Betzios:2017dol,Betzios:2018kwn}, where the imaginary part seems to approach zero as $\ene \to 0$. This indicates that the low-energy plasma in the theory with $\gamma=\gamma_c$ develops quasi-particles. An additional interesting observation is the fact that, for $\gamma=2/10, 3/10$, the low-lying modes exhibit crossings and recombinations --- see \cite{Grozdanov:2019uhi} for a related discussion. 

As mentioned in the main text, QNMs help understand the hydrodynamization of the system. Indeed, for this to happen the relaxation rate of QNMs, which is set by the imaginary part of their frequency \cite{Kovtun:2005ev}, must be larger than the expansion rate of the fluid, which is set by the ratio $\ene'(\tau)/\ene(\tau)$. The time-dependence of the energy density at late times follows from solving  \eqref{eq:equationhydro}  with the replacement $\plong(\tau) \to \peq(\ene(\tau))$, and it takes the form  \cite{Gursoy:2015nza}
\begin{equation}
\label{Elate}
    \ene \propto \tau^{-4/3 + 2\gamma^2} \,.
\end{equation}
This equation, together with the temperature scaling \eqref{Tlate} and the fact that $\omega \sim T$, implies that at late times 
\begin{equation}
\label{relax}
    \left[\frac{\ene'(\tau)}{\ene(\tau)}\right]^{-1}
    \mbox{Im} \, \omega(\ene(\tau))\propto \tau ^{\frac{2+6\gamma^2}{3}}\,.
\end{equation}
Since the exponent is positive we conclude  that the fluid relaxes faster than it expands. The value $\gamma = \gamma_c$ is special because in this case $\mbox{Im} \, \omega \to 0$ as $\ene \to 0$ despite the fact that $T$ approaches a finite value. This means that QNMs take longer to relax in this limit, potentially  jeopardizing the approach to equilibrium \cite{Betzios:2017dol,Betzios:2018kwn}, as in e.g.~\cite{Ecker:2021cvz}. However, even in this case a numerical evaluation of the left-hand side of \eqref{relax}, as displayed in Fig.~\ref{fig:expansion}, shows that this possibility is not realized because the expansion is sufficiently slow. This explains why the system  reaches local equilibrium even in the case $\gamma=\gamma_c$. 

\begin{figure}[t!]
    \centering
    \includegraphics[width=0.49\textwidth]{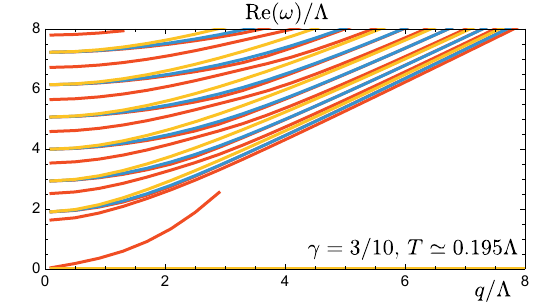}\hfill \includegraphics[width=0.49\textwidth]{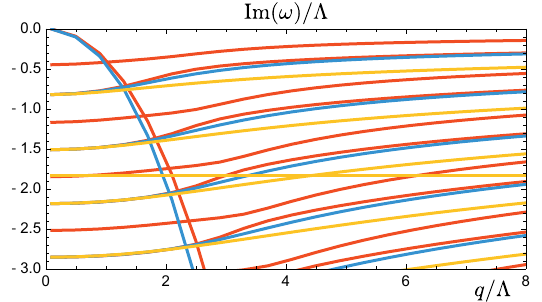}
    \caption{\small Real and imaginary part of QNMs with finite momentum for the spin 2 (yellow), spin 1 (blue) and spin 0 (red) channels. All the finite momentum modes that we have computed behave qualitatively in the same way.}
    \label{fig:QNMq}
\end{figure}

Finally, in order to analyze the behaviour of symmetry-breaking perturbations, we turn our attention to QNMs with non-zero momentum, for which the degeneracy between the different equations disappears. 
Refs.~\cite{Betzios:2017dol,Betzios:2018kwn} analysed perturbations of spacetimes that correspond to our  late-time solutions and no instability was found. Our results for the $q$-dependence of the  QNMs in different regions of parameter space leads to the same conclusion. For example, Fig.~\ref{fig:QNMq}
shows the QNMs frequencies for specific values of $\gamma$ and $T$. We see that no QNM develops a positive imaginary part. 

\end{document}